\documentclass[aps,prb,amsmath,amssymb,twocolumn]{revtex4-2}
\usepackage{graphicx}
\usepackage{bm}
\usepackage{mathptmx}
\usepackage{epstopdf}
\usepackage{booktabs}
\usepackage{multirow}
\usepackage{hhline}
\usepackage{subfigure}
\usepackage{amsmath}
\usepackage{mathrsfs}
\begin{document}
\newcommand{\ms}{m_*}
\newcommand{\bor}{{\bf r}}
\newcommand{\mm}{{\bf m}}
\newcommand{\rdot}{\dot{\bf r}}
\newcommand{\rddot}{\dot\dot{\bf r}}
\newcommand{\es}{\bf{E^s_\sigma}}
\title{Magnetization dynamics in skyrmions due to high-speed carrier injections from Dirac half-metals}
\author{Satadeep Bhattacharjee$^{1}$}
\email{s.bhattacharjee@ikst.res.in}
\affiliation{Indo-Korea Science and Technology Center (IKST), Bangalore, India}
\author{Seung-Cheol Lee$^2$}
\affiliation{Electronic Materials Research Center, Korea Institute of Science $\&$ Technology, Korea}
\begin{abstract}
Recent developments in the magnetization dynamics in spin textures, particularly skyrmions, offer promising new directions for magnetic storage technologies and spintronics. Skyrmions, characterized by their topological protection and efficient mobility at low current density, are increasingly recognized for their potential applications in next-generation logic and memory devices. This study investigates the dynamics of skyrmion magnetization, focusing on the manipulation of their topological states as a basis for bitwise data storage through a modified Landau-Lifshitz-Gilbert equation (LLG). We introduce spin-polarized electrons from a topological ferromagnet that induce an electric dipole moment that interacts with the electric gauge field within the skyrmion domain. This interaction creates an effective magnetic field that results in a torque that can dynamically change the topological state of the skyrmion. In particular, we show that these torques can selectively destroy and create skyrmions, effectively writing and erasing bits, highlighting the potential of using controlled electron injection for robust and scalable skyrmion-based data storage solutions.
\end{abstract}
\keywords{Toplogical ferromagnet, quantum magnet, skyrmion, magnetic memory}
\maketitle

\section{Introduction}
The field of magnetization dynamics in magnetic textures such as skyrmions and vortices has seen significant progress in recent years~\cite{text1,text2,antos2008magnetic}. Magnetic skyrmions, which are quasiparticle-like textures with a unique topology, have attracted attention because of their potential applications in spintronics and data storage~\cite{sk1,sk2,sk3,Fert2013SkyrmionsOT}. Recent research has focused on developing efficient methods to create, manipulate, and excite magnetic skyrmions~\cite{zhang2023,yang2024acoustic,hu2017creation}. It has been theoretically shown by Koshibae \textit{et al.}~\cite{koshibae2014creation} that even local heating can produce magnetic skyrmions in chiral and magnetic dipole magnets.
Other topological magnetic textures, such as domain walls, vortices, and bimerons, have also been studied. The interaction between spin waves and magnetic textures has been found to be particularly interesting and rich in physics, leading to the development of magnon-based information processing schemes. The flexibility and reconfigurability of magnetic textures have been discussed regarding their potential for applications in various fields. The recent progress in the field of magnetization dynamics in magnetic textures therefore has opened up new avenues for the development of practical devices with improved performance.

Skyrmions, topologically protected spin textures, have become a focal point in magnetism research due to their potential applications in spintronic devices. Their unique properties, combined with the ability to be moved with low current densities, make them promising candidates for next-generation memory and logic devices. 
Current-induced destruction of skyrmions is a phenomenon that occurs when a skyrmion encounters a critical current density, at which it experiences a force that causes it to annihilate. This process is particularly useful in the context of spintronics, where the manipulation and control of skyrmions are essential for various applications, such as memory devices and logic circuits.
With the increased interest in designing novel spintronic devices using quantum materials, it is naturally interesting to study the topological materials/skyrmion supporting ferromagnet heterostructures~\cite{pesin2012spintronics,baker2015spin,jonietz2010spin,manchon2015new,Chen2019Evidence,Dahir2018Interaction,Zarzuela2019Stability}. Heterostructures combining ferromagnets with topological insulators have shown promise for hosting magnetic skyrmions, offering a novel platform for spintronics and information storage technologies~\cite{Dou2023Deterministic,Saini2023Spin}. Studies have demonstrated the presence of magnetic skyrmions and skyrmionium at the interface of these heterostructures, facilitated by strong spin-orbit coupling and interfacial Dzyaloshinskii–Moriya interaction, which are key for the development of next-generation low-energy spintronic devices~\cite{Li2019Micromagnetic}. 

In this paper, we explore the magnetization dynamics of skyrmions, with an emphasis on manipulating its topological behavior, using a modified Landau-Lifshitz-Gilbert (LLG) equation~\cite{gilbert2004phenomenological}, particularly, we proposed that if spin-polarized electrons are incident from a topological ferromagnet such as Dirac half metal (DHM), then these stream of electrons with relativistic speed can give rise to an electric dipole moment which can interact with gauge electric field in the region of a spin texture such as skyrmion. Such interaction can give rise to an effective magnetic field that can offer an additional torque to the LLG equation similar to the spin-orbit torques~\cite{mahfouzi2016antidamping,haidar2019single}. DHM are materials that exhibit unique electronic properties, combining the characteristics of half-metals~\cite{bhattacharjee2019first} and Dirac fermions~\cite{chen2022}. These materials are of significant interest in the field of spintronics due to their potential for high-speed spintronics applications with zero energy consumption. The electronic structure of DHMs is characterized by a band structure with a gap in one spin channel and a Dirac cone in the other spin channel. This means that one spin channel has a bandgap, while the other spin channel has a linear dispersion relation, similar to that of massless Dirac fermions. This unique electronic structure is responsible for the spin-polarized transport properties of DHMs~\cite{wang2018high}. Also, recently it has been shown that such interaction can give rise to additional anomalous velocity that would give rise to transverse conductivity similar to the anomalous Hall effect~\cite{bhattacharjee2023spin}. In this work, we concentrate on how this new kind of spin torque can affect the magnetization dynamics in a typical spin texture such as skyrmion. It is well known that skyrmion is a topologically stable magnetic vortex that has been proposed as a promising candidate for next-generation information storage and processing devices~\cite{buttner2018theory}. We show that the above mentioned type of spin torque can be used to create and destroy skyrmions and that the concentration, direction, and velocity of the incident carriers can control skyrmion dynamics. It should be noted that one of the key advantages of using skyrmions in spintronics devices is their potential for low energy consumption. The ability to move skyrmions with currents several orders of magnitude lower than those required for typical domain wall motion could lead to more energy-efficient devices. However, this also means that understanding the destruction mechanisms caused by current injection is crucial to prevent accidental data loss. For simplicity, here we consider a single skyrmion for study. This particular work addresses the treatment of the interface between quantum materials such as DHM and skyrmion from a new perspective. We discuss the results after presenting the theoretical framework we used.

\section{Theoretical framework: the effective field and torque due to the interaction between the gauge electric field and the dipole moment of the }
\begin{figure}
    \centering
    \includegraphics[width=0.85\linewidth]{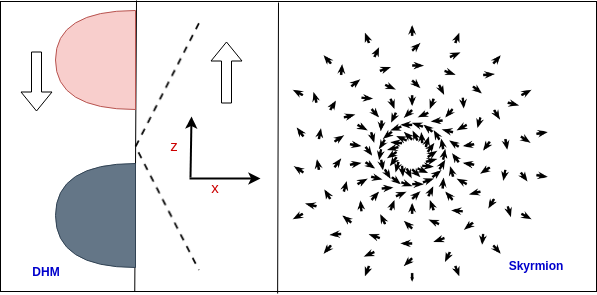}
    \caption{Schematic representation of the interface between the Dirac Half metal (DHM) and the skyrmion. The big arrows on the left-hand side represent the spin-up and spin-down channels.}
    \label{Schematic}
\end{figure}

The theoretical framework is formulated by emphasizing the significance of the interplay between the gauge field and the dipole moment in influencing the behavior of magnetic systems. In this section, we review the mathematical formulations and the physical implications of this interaction in the context of skyrmions. The geometry of the problem is shown in the Fig.\ref{Schematic}. 
In the present investigation, we consider a planar geometry confined to the xz-plane, where a skyrmion is localized within a thin film structure at the interface with a Dirac Half-metal (DHM)~\cite{liu2017yn,ma2018rhombohedral,ishizuka2012dirac}. The skyrmion exhibits a non-trivial spin configuration where the magnetization vector \(\mathbf{M}\) wraps around a unit sphere. The DHM injects a stream of spin-polarized electrons, their polarization modulated by the relativistic band structure of the material. The interactions between these electrons and the skyrmion occur predominantly along the interface in the xz-plane, influencing the skyrmion's dynamic response to external perturbations and the resultant magnetization dynamics, governed by the skyrmion's complex geometry within this plane. Suppose the skyrmion spin texture is described by the magnetic vector $\mathbf{M}$. We are interested in the dynamics of it as we inject a stream of spin-polarized electrons from a topological ferromagnet in the vicinity of it. Building on our previous work, we now turn our attention to the torque and effective field arising from the continuity equation of the magnetization density. A comprehensive analytical and numerical analysis will be presented to elucidate the dynamics of skyrmions under various scenarios. The current-induced magnetization dynamics is described by the modified LLG equation~\cite{bhattacharjee2023spin},
\begin{equation}
\dot{\bf M} = \gamma {\bf M} \times {\bf H_{eff}} + \alpha {\bf M} \times \dot{\bf M} + \gamma{\bf T^{\text{gauge}}}
\label{LLG}
\end{equation}
Where the first two terms in the right hand of the above equation describe the usual LLG equation in terms of precessional and damping torque. The last term $\mathbf{T}^{\text{gauge}}$ is the torque due to the injection of spin-polarized carriers from the Dirac half-metal. This torque arises due to the interaction of the very fast carriers with velocity $\mathbf{v}$ and magnetization density $\mathbf{m}$ near the interface with the gauge electric field in the skyrmion.  Because of the relativistic nature of these electrons, there is an electric dipole moment given by, ${\bf d}=\frac{1}{c^2}({\bf v}\times {\bf m})$~\cite{panofsky2005classical,vekstein2011comment,griffiths2013mansuripur}. Here $c$ is the velocity of the light. The magnetization density can be split into two parts as $\mathbf{m} = \mathbf{m}_0 + \delta \mathbf{m}$~\cite{matos2009spin}, where $\mathbf{m}_0$ is the equilibrium magnetization density of the conduction electrons in the skyrmion region without any injected carriers (in the absence of DHM), and $\delta \mathbf{m}$ is the modification of it due to carrier injection from DHM.
The above torque is a result of the interaction of the gauge electric field ($\mathbf{E^s}$) and the above-mentioned dipole via,
\begin{equation}
\mathscr{E}=-\mathbf{E}^s \cdot \mathbf{d}=-\sum \frac{\hbar}{2 e} \mathbf{M} \cdot\left(\dot{\mathbf{M}} \times \nabla_i \mathbf{M}\right) d_i
\label{int}
\end{equation}
The $i^{th}$-component of the gauge field can be written as $E_i^s= \pm \frac{\hbar}{2 e} \mathbf{M} \cdot\left(\dot{\mathbf{M}} \times \nabla_i \mathbf{M}\right)$~\cite{tatara2019effective,araki2020magnetic,jalil2014topological}. The sign $\pm$ refers to the spin of the carrier (either up or down) experiencing the field. 
Here $\hbar$ is the reduced Planck constant, $e$ is the elementary charge respectively.
The effective field $\mathbf{H}_{eff}$ in Eq.\ref{LLG} is derived from the background magnetic structure which follows a Dzyaloshinskii–Moriya interaction (DMI) given by~\cite{koretsune2015control},
\begin{equation}
E_{\text{DMI}} = \int \mathbf{D} \cdot (\mathbf{M} \cdot (\nabla \times \mathbf{M})) \, d^3 r
\end{equation}
The effective field is given by ${\bf H}_{\text{eff}} = -\frac{1}{\mu_0 M_s} \frac{\delta E_{\text{DMI}}}{\delta\mathbf{M}}$
It was shown in the previous study~\cite{bhattacharjee2023spin} that the nature of the new damping term depends on the spin-polarization direction of the incident electrons. When the incident electrons are spin-polarized along the z-direction $[m = (0, 0, m_z)]$, the additional damping acts opposite to the intrinsic damping and lowers the effective damping in the system. Therefore, this torque is a kind of anti-damping torque that tries to amplify the processional motion. On the other hand, if the incident electrons are polarized along the y-direction $[m = (0, m_y, 0)]$, it has tendency to enhance the total damping in the system. Therefore, the above interactions give both damping-like and anti-damping-like contributions depending on the direction of the polarization of the incident electrons. 
\noindent The entire theoretical formalism relies on two assumptions:
\begin{enumerate}
\item Equivalence of drift and Fermi velocities, as observed in most of the topological materials~\cite{shishir2009velocity}, enabling the formation of electric dipole moment (\(\mathbf{d}\)) to interact with the gauge electric field (\(\mathbf{E}_s\)).
\item The gauge electric field and the electric dipole moment due to fast carriers are in the same coordinate system, permitting their dot product, \( -\mathbf{E}_s \cdot \mathbf{d}\), to represent interaction energy.
\end{enumerate} 
The first assumption comes from the fact that for materials with linear band structures, like graphene, the drift velocity is comparable to the Fermi velocity, indicating that similar conditions are expected in Dirac half metals (DHMs) and other topological materials which are considered as carrier injector to the spin-texture here. To fulfill the second assumption, we can perform a coordinate transformation such that  \(\mathbf{E}_s\) and \(\mathbf{d}\) are in the same reference frame.  This involves rotating \(\mathbf{d}\) using a suitable rotation matrix.

\section{Rotation matrix to align the dipole vector with the texture frame of reference}

The rotation matrix, denoted here as \(R\), is essential for aligning vectors from a global reference frame to the local frame defined by a skyrmion's magnetization texture at any given point. Given its non-uniform and topologically non-trivial magnetization profile, this alignment is crucial for accurately analyzing the interactions and dynamics within the skyrmion structure. In our scenario, the dipole orientation is either aligned with the global \(\mathbf{y}\) axis or with the global \(\mathbf{z}\) axis because the incident carriers are injected along the global \(\mathbf{x}\) direction. 

Given a local magnetization direction \(\mathbf{M} = [M_x, M_y, M_z]\) at a point within the skyrmion, and the global z-axis \(\mathbf{z} = [0, 0, 1]\), for example, the construction of \(R\) involves the following steps:

\begin{enumerate}
    \item \textbf{Determine the Rotation Axis:} The axis around which the rotation occurs, \(\mathbf{k}\), is found by taking the cross product of \(\mathbf{z}\) and \(\mathbf{M}\), and normalizing the result:
    \[\mathbf{k} = \frac{\mathbf{z} \times \mathbf{M}}{\|\mathbf{z} \times \mathbf{M}\|}.\]

    \item \textbf{Calculate the Rotation Angle:} The angle \(\theta\) of rotation is determined by the angle between \(\mathbf{z}\) and \(\mathbf{M}\), computed using the dot product:
    \[\theta = \arccos\left(\frac{\mathbf{z} \cdot \mathbf{M}}{\|\mathbf{z}\| \|\mathbf{M}\|}\right).\]

    \item \textbf{Construct the Skew-Symmetric Matrix:} The skew-symmetric matrix \(K\) associated with the rotation axis \(\mathbf{k}\) is constructed as follows:
    \[K = \begin{bmatrix}
    0 & -k_z & k_y \\
    k_z & 0 & -k_x \\
    -k_y & k_x & 0
    \end{bmatrix}.\]

    \item \textbf{Apply Rodrigues' Rotation Formula:} The rotation matrix \(R\) is then derived using Rodrigues' rotation formula, which in terms of \(K\) and \(\theta\), is given by~\cite{dai2015euler}:
    \[R = I + \sin(\theta) K + (1 - \cos(\theta)) K^2,\]
    where \(I\) is the identity matrix.
\end{enumerate}

This formulation of \(R\) allows for the transformation of any vector from the global reference frame to the local frame at the point of interest within the skyrmion. The spatial variation of \(\mathbf{M}\) across the skyrmion necessitates the computation of \(R\) at each point where an alignment with the local magnetization direction is required. To ensure that both \(\mathbf{E}_s\) and \(\mathbf{d}\) are in the same reference frame, the following transformation is performed,
\begin{equation}
\mathbf{d}'=R\mathbf{d}
\end{equation}

\section{Interaction Energy and Torque Calculation}
\label{sec:interaction}

The interaction energy \(\mathscr{E}\) is defined as,
\begin{equation}
\mathscr{E}=-\mathbf{E}^s \cdot \mathbf{d'}=-\sum \frac{\hbar}{2 e} \mathbf{M} \cdot\left(\dot{\mathbf{M}} \times \nabla_i \mathbf{M}\right) d_i'.
\end{equation}
As explained above, this interaction is in general \textit{local} and depends on where it takes place in the xz-plane.
The torque \(\mathbf{T}^{\text{gauge}}\), attributable to the interaction, follows:
\begin{equation}
\mathbf{T}^{\text{gauge}} = \mathbf{M} \times \mathbf{H}^{\text{gauge}},
\end{equation}
with \(\mathbf{H}^{\text{gauge}}\) being:
\begin{equation}
\mathbf{H}^{\text{gauge}}=-\frac{\partial \mathscr{E}}{\partial \mathbf{M}}=\frac{\hbar}{2 e} \sum_i\left(\dot{\mathbf{M}} \times \nabla_i \mathbf{M}\right) d_i'.
\end{equation}
\section{Numerical Implementation and Analysis}
\label{sec:numerical}

For the numerical simulation of magnetization dynamics within skyrmions, we used the physical constants: gyromagnetic ratio \(\gamma\)=1.7608597$\times 10^{11}$ rad/(sT), Gilbert damping coefficient \(\alpha\)=0.01, reduced Planck constant \(\hbar\)=1.054571834$\times 10^{-34}$ Js, elementary charge \(e\)=1.60217663$\times 10^{-19}$ C, speed of light \(c\)=3$\times 10^8$ m/s. The drift velocity \(v\) of the carriers was varied in the simulation. As already mentioned, the effective magnetic field \(\mathbf{H}_{\text{eff}}\) is defined mainly in terms of the Dzyaloshinskii–Moriya interaction. The DMI constant used in this work was D=0.1 meV/\AA. Other than DMI, a small field of 0.1T was added in the \(\mathbf{H}_{\text{eff}}\) to take account of the other internal fields.
The temporal resolution is established to proceed through 100 discrete time steps.

\begin{figure}
    \centering
    \includegraphics[width=0.5\textwidth]{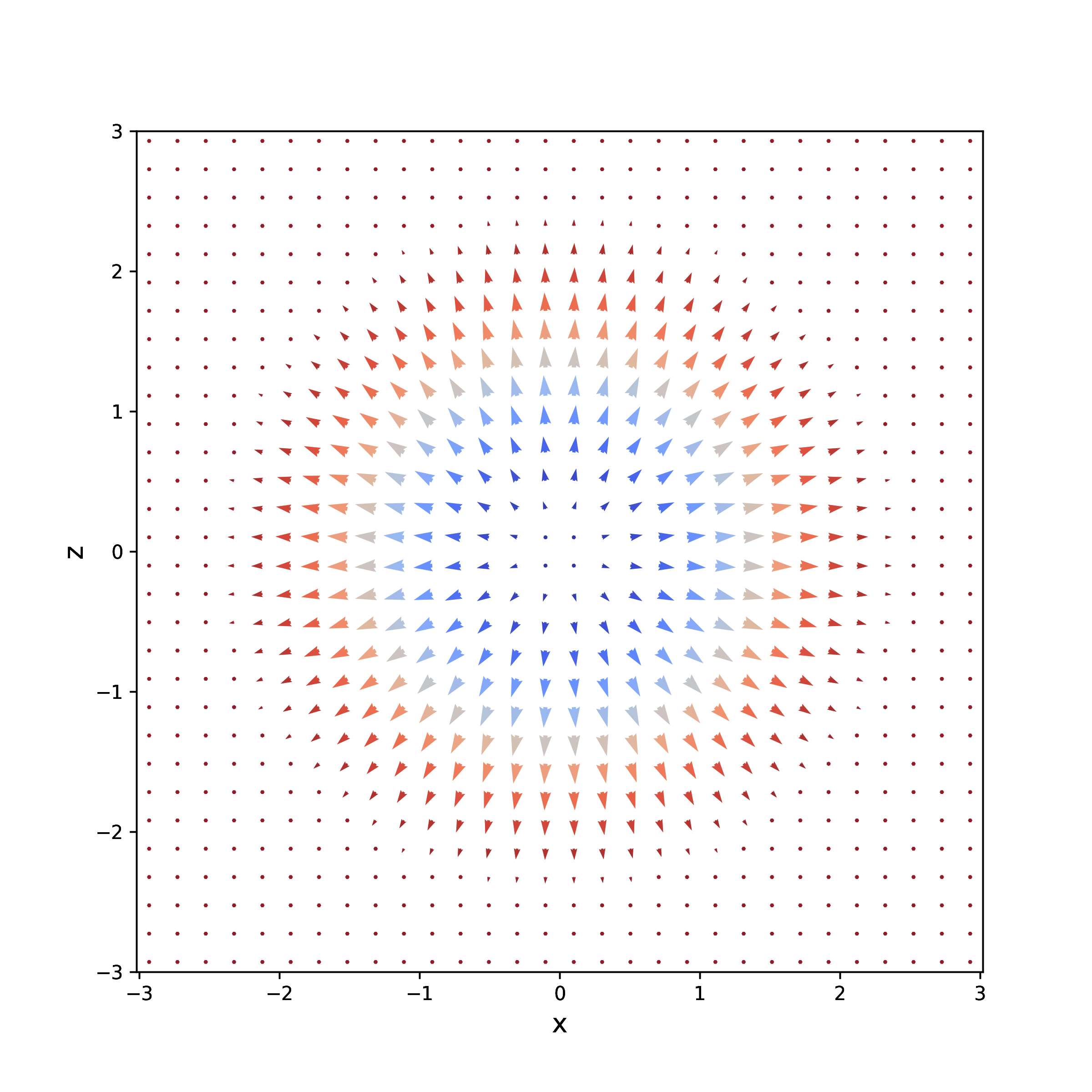}
    \caption{Initial skyrmion profile targeted for spin injection from the Dirac Half-metal. The skyrmion profile is shown in the global xz plane.}
    \label{fig:Initial}
\end{figure}

The skyrmion profile can be visualized within a two-dimensional lattice, as depicted in Fig.~\ref{fig:Initial}, which acts as a precursor for the spin injection from the adjacent topological ferromagnet, specifically a Dirac Half-metal.
In the Figure, it can be seen that the spins are pointing radially outward from the center of the skyrmion, which is a characteristic feature of Neel skyrmions. 
The skyrmion's out-of-plane magnetization component \(M_y\) is described as a radially dependent function, while its in-plane components \(M_x, M_z\) exhibit a vortical arrangement. We adopt an advanced midpoint method to iteratively update the magnetization vectors, ensuring the incorporation of both damping effects and the spin torque influences. At each time step, we calculate the torque exerted on the skyrmion to predict its evolution.

To quantify the resulting magnetic configurations, we calculate the topological charge \(Q\) of the skyrmion, a conserved quantity emblematic of the skyrmion's stability and defined as~\cite{rybakov2019chiral,Jalil2014TopologicalHC,zhao2022spin},
\begin{equation}
Q = \frac{1}{4\pi} \int (\mathbf{M} \cdot (\partial_x \mathbf{M} \times \partial_z \mathbf{M})) \, dx \, dz,
\label{eq:topological_charge}
\end{equation}
where \(\partial_x \mathbf{M}\) and \(\partial_z \mathbf{M}\) are the spatial derivatives of the magnetization vector \(\mathbf{M}\) along the \(x\) and \(z\) directions, respectively. Here, \( Q = 1 \) indicates a stable, topologically non-trivial state such as a skyrmion, where the magnetization wraps around the sphere once. On the other hand, \( Q = 0 \) indicates either a trivial state or a state where any local twists or curls in the magnetization do not contribute to a net topological charge, suggesting a topologically trivial state. The dot product \(\mathbf{M} \cdot (\partial_x \mathbf{M} \times \partial_z \mathbf{M})\) quantifies the local contribution to this topological winding. This integral is numerically approximated over the simulation grid.
\section{Results and discussions}
Skyrmions, as topologically protected magnetic textures, hold great promise as information carriers in the field of spintronics. Their stability and manipulability through external fields and currents make them excellent candidates for the next generation of memory and logic devices. Our investigation into the magnetization dynamics of skyrmions focuses on the primarily interplay between the Dzyaloshinskii-Moriya (DM) interaction and the interaction between the relativistic dipole moment and the gauge field.
It should be noted that rotation matrix \(R\), which encapsulates the transformation required to align vectors from a global reference frame to the local magnetization direction of a skyrmion, exhibits significant spatial variation across different positions within the skyrmion structure. This variation is a direct reflection of the skyrmion's unique magnetization texture, which transitions from a specific orientation at the edge, often tangentially arranged around the skyrmion boundary to a markedly different orientation at the center, where the magnetization typically aligns either parallel or antiparallel to the skyrmion's axis. 
To consider a realistic situation let us consider that \( m(\mathbf{r}) \) has its maximum value at the interface between the DHM \& skyrmion. It decays exponentially from the edge to the core. Such exponential decay of \( m(\mathbf{r}) \) can be modeled as \( m(\mathbf{r}) = m_0 \exp(-\lambda r) \), where \( m_0 \) is the maximum magnetization density at the edge, \( \lambda \) is a decay constant, and \( r \) is the radial distance from the edge towards the skyrmion core. Such dependence on injected carriers' influence on the skyrmion structure is strongest at the periphery and weakens as one moves toward the center. 
Considering such a profile of the \( m(\mathbf{r}) \), let us consider a point (x,z) which lies almost outside the range of the skyrmion radius, where $\theta\sim 0$, there \( \mathbf{M} \) almost aligns with the z-axis, R is an identity matrix, and therefore $\mathbf{d}'\sim\mathbf{d}$. This situation is equivalent to the original work of Bhattacharjee \textit{et al.}\cite{bhattacharjee2023spin}.
To simulate a realistic environment for skyrmion stability, an additional magnetic field of 0.1 Tesla is applied, because skyrmion stability is compromised at fields greater than 1.0 Tesla, even in the absence of any interactions with a spin gauge field. The DM interaction is taken as the predominant contribution to the effective magnetic field $\mathbf{H}_{eff}$, which governs the behavior of skyrmions in the system. Figure \ref{fig:Initial} illustrates the initial configuration of the skyrmion with a topological charge $Q = 1$, representative of an isolated skyrmion state that is stabilized by the interplay between the DM interaction and the external magnetic field.

As we introduce fast-moving carriers into the system, the magnetization dynamics in the skyrmion are drastically altered. These carriers are modeled as spin-polarized electrons originating from a DHM, chosen for their high spin polarization and relativistic effects, which are significant in interacting with the skyrmion's internal structure.

\begin{figure}
\includegraphics[width=1.0\columnwidth]{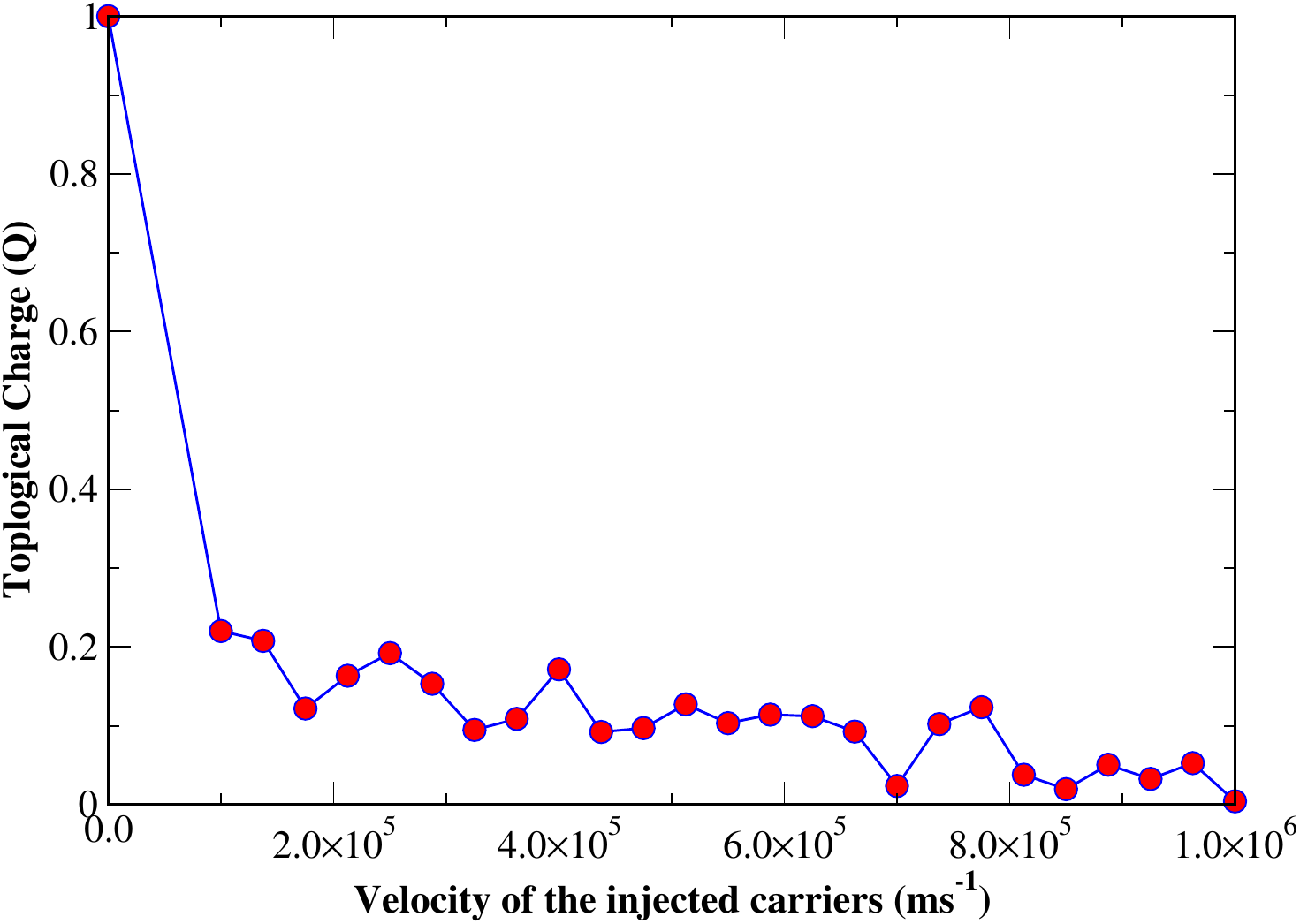}
\caption{Dependence of topological charge $Q$ on carrier velocity $v$ in a skyrmion subjected to spin-polarized electrons from a Dirac Half Metal.}
\label{velocity}
\end{figure}

The carrier injection effect is illustrated in Figure \ref{velocity}. Here we consider the case when carriers injected from the DHM have the spin polarization along z-direction, i,e $\mathbf{m}=(0,0,m_z)$. As we have shown in our previous study such a situation will create an anti-damping sort of spin torque~\cite{bhattacharjee2023spin}. (It can also be seen easily here that for geometry we consider, the effective magnetic field due to the interaction between the gauge field and dipole moment is given by \(
\mathbf{H}^{\text{gauge}} \sim \frac{\hbar}{2e} \left(\dot{\mathbf{M}} \times \nabla_y \mathbf{M}\right) \frac{1}{c^2} v_x m_z
\) which implies that the torque, \(
\mathbf{T}^{\text{gauge}} = \mathbf{M} \times \mathbf{H}^{\text{gauge}}
\) is perpendicular to \(\mathbf{M}\)). 
The plot demonstrates a nonlinear correlation between the velocity of carriers and the skyrmion's topological charge $Q$ as introduced in Eq.\ref{eq:topological_charge}.
The results correspond to the solution of the modified LLG equation. The simulation employs a time-evolution method to study the dynamics of a skyrmion within a magnetized material, incorporating both classical and quantum mechanical effects. The evolution of the magnetization vector, \(\mathbf{M}\), is governed by the LLG equation, modified to include an additional term accounting for the gauge torque arising from the relativistic interaction between the magnetization dynamics and a moving electron current. The effective field, \(\mathbf{H}_\text{eff}\), combines the additional magnetic field and the Dzyaloshinskii-Moriya Interaction (DMI), which is essential for stabilizing the skyrmion structure. The simulation discretizes time into small steps of 1 picosecond (ps), using a midpoint method for numerical integration to ensure stability and accuracy in the evolution of the magnetization. The total simulation time is 100 ps. Topological charges are calculated at specified intervals to monitor the skyrmion's behavior and integrity throughout the simulation, offering insights into its dynamic properties and interactions with external influences.
Fig.\ref{velocity} depicts a plot of the topological charge (Q) as a function of carrier velocity (v).  As mentioned, with no carrier injection, the topological charge Q=1. As the velocity increases, the topological charge rapidly decreases, dropping to approximately 0.2 at \(2.0 \times 10^5\) ms\(^{-1}\), and further diminishing to a value of 0.004 at a velocity of \(1.0 \times 10^6\) ms\(^{-1}\). The plot corresponds to the value $m_0=10^{18}$ $m^{-3}$ and $\lambda=1$.
This trend suggests that the skyrmion's stability is affected by the carrier velocity; higher velocities lead to a reduction in topological charge. The topological charge quantifies the degree to which the skyrmion's spin texture wraps around the unit sphere. A charge close to 1 implies a stable, well-defined skyrmion, while lower values indicate a distortion or unwinding of the skyrmion structure, possibly due to the increased influence of the relativistic torque at higher electron velocities. This could be interpreted as the skyrmion structure being less robust and potentially more prone to deformation or annihilation when subjected to faster moving electric carriers, which would be consistent with a scenario where the interplay between the spin texture and the itinerant electrons is influenced by relativistic effects such as spin-transfer torque.

In the simulation of skyrmion dynamics under the influence of a current-induced spin-transfer torque, a crucial parameter is the critical current density, beyond which the skyrmion becomes unstable and potentially annihilates. For the above case, the skyrmion is stable with a topological charge \( Q \) is \( 0.99\) at a carrier velocity of \( 0.025 \times 10^6 \) m/s.  We may infer that the skyrmion configuration remains nearly intact, as \( Q \) is very close to 1, the value expected for an undistorted skyrmion. Considering this as the highest velocity at which \( Q \) remains close to 1, we can consider it the critical velocity \( v_{\text{crit}} \). With the assumption of 100\% spin polarization, provided by the use of Dirac half-metals, we take the net magnetization density \( m \) as equivalent to the charge carrier density \( n \). Thus, the critical current density \( J_{\text{crit}} \) that marks the boundary of skyrmion stability can be calculated using the relation \( J_{\text{crit}} = n e v_{\text{crit}} \). Substituting \( n = 0.5 \times 10^{18} \) m\(^{-3}\), \( e = 1.602 \times 10^{-19} \) C, and \( v_{\text{crit}} = 0.025 \times 10^6 \) m/s, we obtain a critical current density of approximately \( 2002 \) A/m\(^2\).
\begin{figure}
\includegraphics[width=1.0\columnwidth]{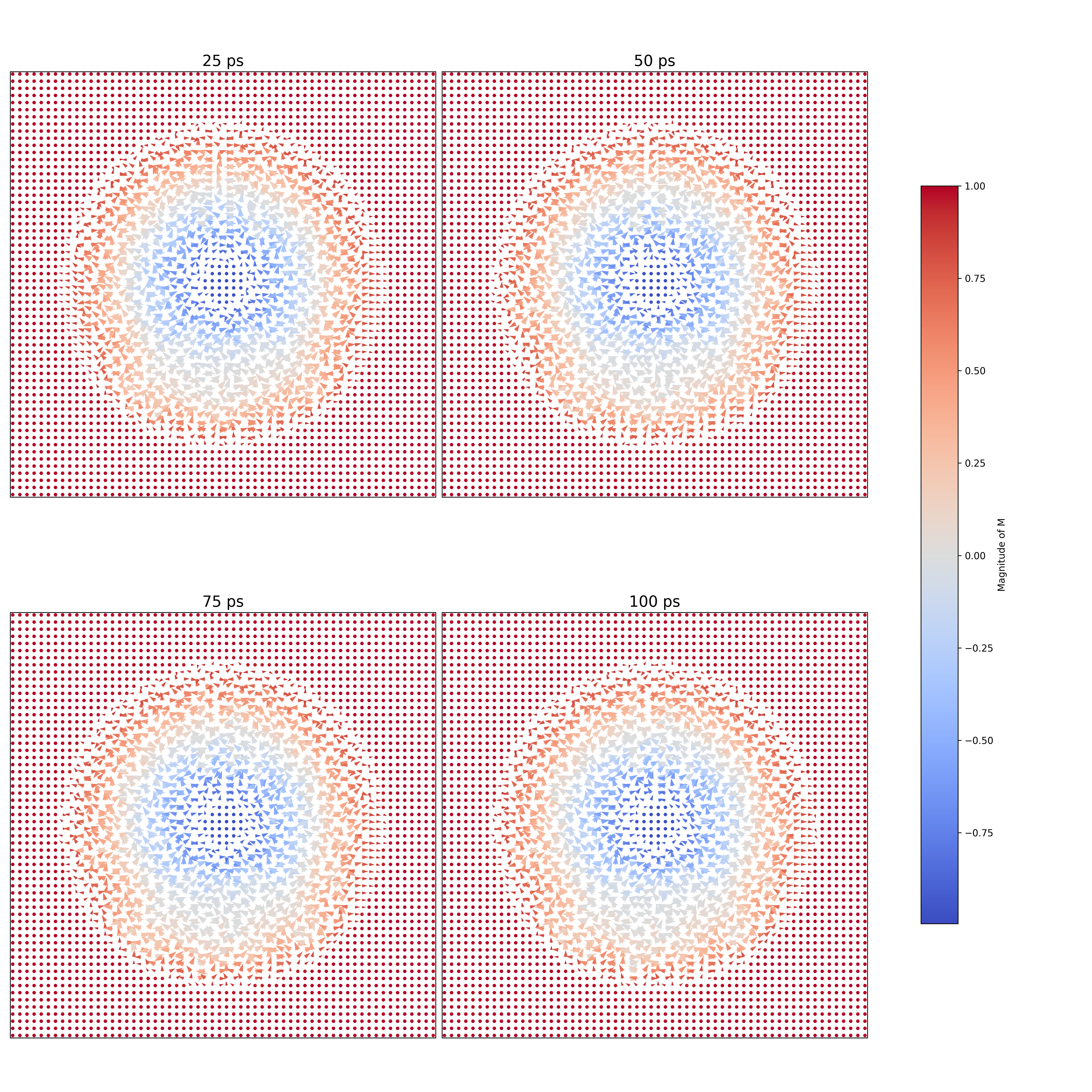}
\caption{Time evolution of the skyrmion when injected with carriers with velocity $v=1.0\times 10^6$ m/s. Snapshots are taken at 25ps, 50ps, 75ps and 100ps. }
\label{Time}
\end{figure}
To understand how the is destroyed at the carrier velocity of $1.0\times 10^6$ ms$^{-1}$, we plot the temporal evolution of the skyrmion in Fig.4. At a carrier injection velocity of \( v = 1.0 \times 10^6 \) m/s, the time-resolved magnetization configurations show that skyrmion stability gradually decreases and eventually the skyrmion is destroyed. This decay in topological charge suggests a weakening of the topological protection due to the increasing influence of spin torque effects. The spin torque, proportional to the carrier velocity in the LLG equation, destabilizes the skyrmion's structure by exerting a non-conservative force on the local magnetic moments. This force opposes the intrinsic stabilization mechanisms, such as the Dzyaloshinskii-Moriya interaction (DMI)  leading to the unwinding of the skyrmion. While the plots in Fig.\ref{Time} show subtle visual changes, the significant reduction in topological charge indicates underlying alterations in the spin texture, possibly including the annihilation of the skyrmion core or the disruption of its chiral boundary.

The plot in Fig.\ref{Time} depict the temporal evolution of a skyrmion subject to spin-polarized electronic currents at a velocity of \( v = 1.0 \times 10^6 \) m/s. The figure illustrates the magnetization dynamics in a skyrmion induced by the injection of spin-polarized electrons at different time frames (25 ps, 50 ps, 75 ps, and 100 ps). The calculated topological charges (Q) for each time frame are 0.11, 0.04, 0.01, and 0.004, respectively. The initial topological charge at 25 ps is already very small, indicating a weakly stable skyrmion from the outset. As the skyrmion experiences the perturbation from the spin-polarized electron injection, this small initial topological charge further decreases, reflecting a rapid loss of the skyrmion's topological protection. This decay in topological charge highlights the significant influence of external spin currents on the stability and eventual dissipation of the skyrmion itself. 

\begin{figure}
\includegraphics[width=1.0\columnwidth]{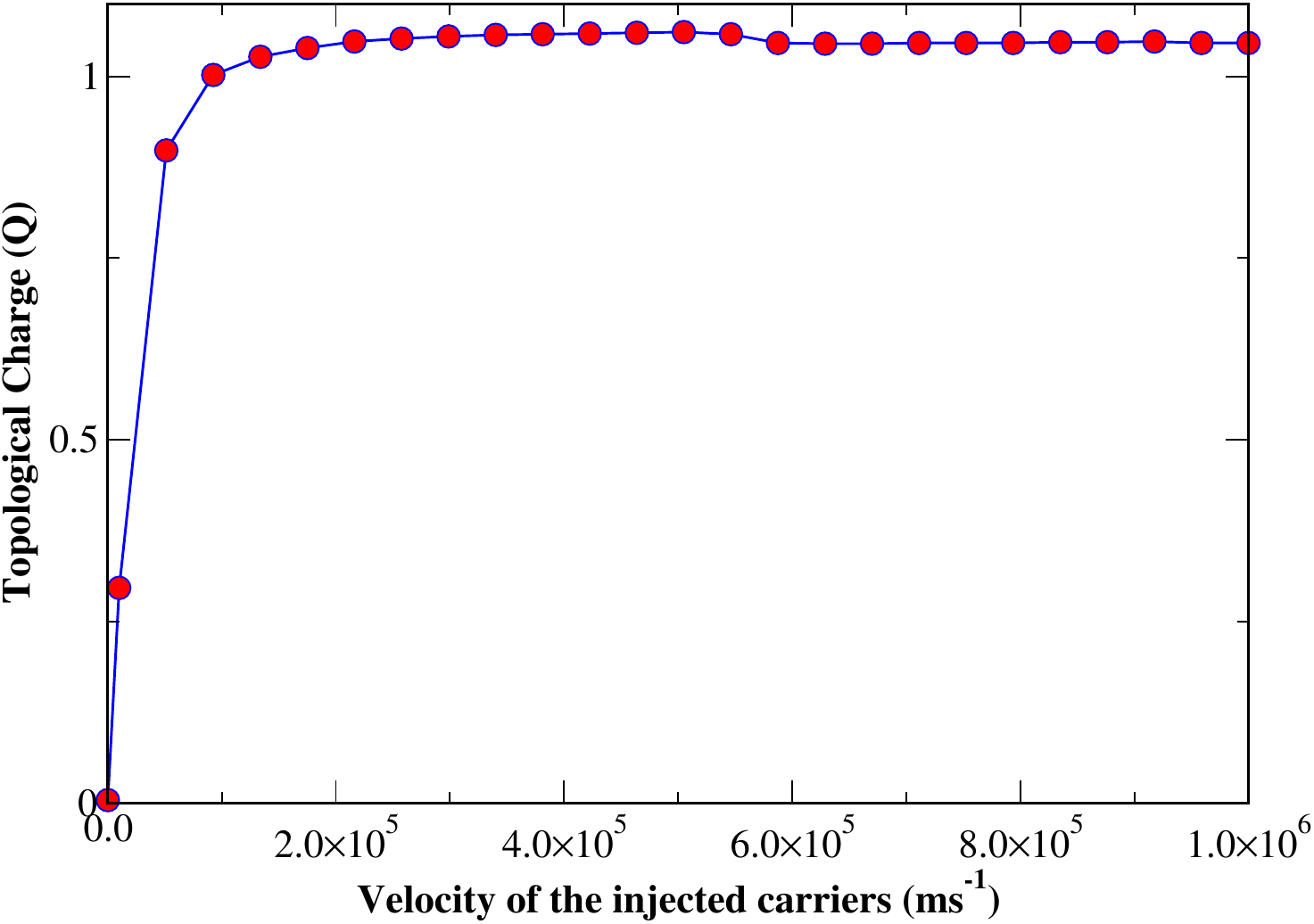}
\caption{Evolution of the topological charge $Q$ for the spin injection to the spin texture with negligible topological charge.}
\label{Fig5}
\end{figure}

Finally, it can be seen that the destroyed skyrmion can restore its topological state if we inject current such that the incident electrons have along the y-direction $[m = (0, m_y, 0)]$. Again from our previous study, this corresponds to the case when the spin torque has a damping-like feature~\cite{bhattacharjee2023spin}. 
To demonstrate this we start with the spin texture with a topological charge of $Q\sim 0.004$~(the final state after 100 ps of dynamics, we now call it a simple spin texture as it has very small Q) as shown in the Fig.\ref{Time}. Fig.\ref{Fig5} shows the evolution of the topological charge, $Q$, with respect to the carrier velocity, $v$, providing significant insight into the dynamic behavior of the texture under current-induced influences. It can be seen that due to numerical error, the topological charge is slightly overestimated.  This plateau hints at a potential asymptotic approach to the skyrmion's initial state or a dynamic equilibrium established by the interplay between the external current injection and the inherent material properties and internal fields. 

Therefore, in our study, the anti-damping-like torque likely provided sufficient energy to misalign the spins within the skyrmion structure, leading to its destruction due to the destabilization of its coherent spin structure. It should be noted here that Litzius \textit{et al.} have experimentally found that a field like  SOT can change the skyrmion Hall angle with increasing current density~\cite{litzius2017skyrmion}. 
Conversely, the damping-like torque aided in realigning the spins in a way that either supported the creation or the stabilization of the skyrmion.  
\begin{figure}
\includegraphics[width=1.1\columnwidth]{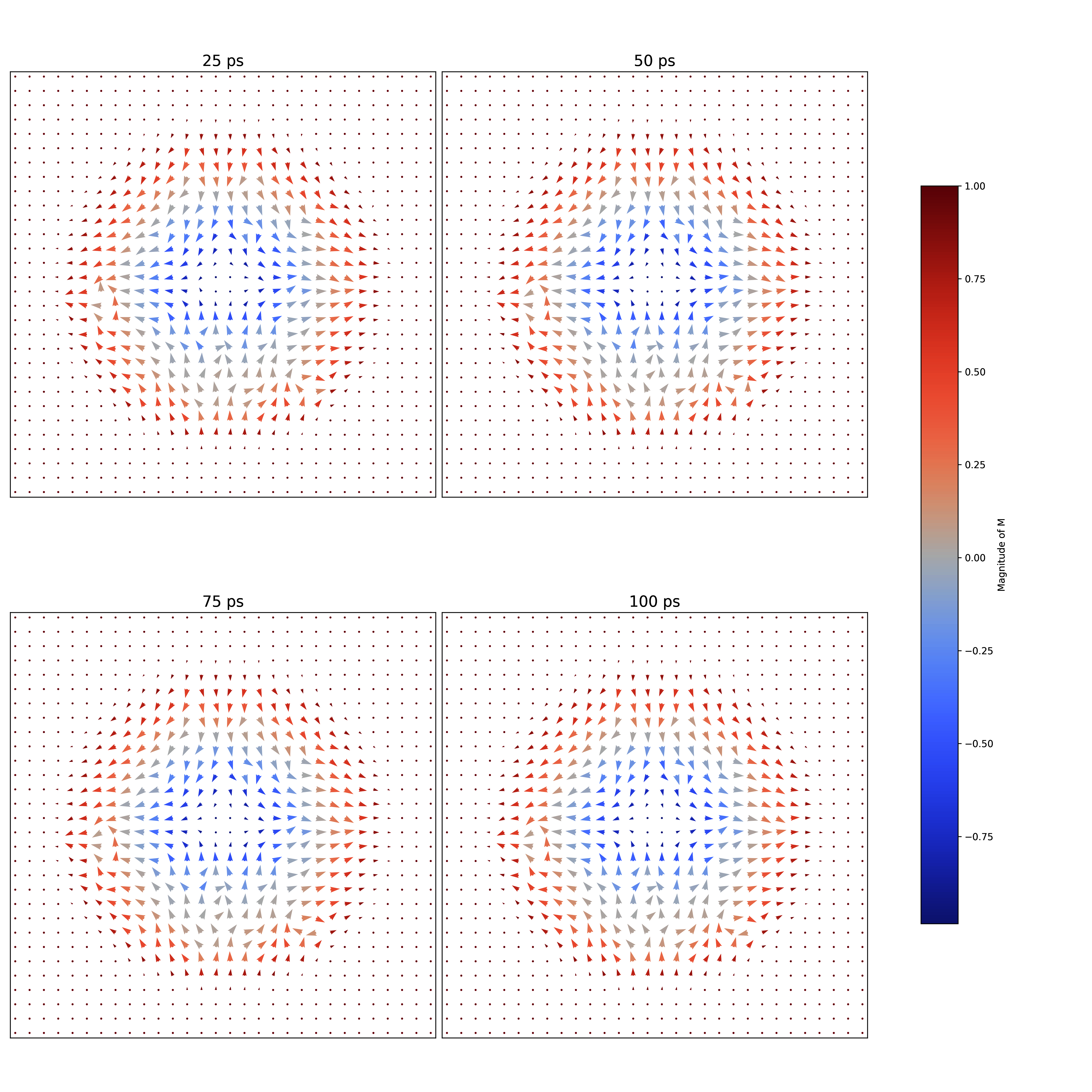}
\caption{Reacreation of the skyrmion when injected with carriers with velocity $v=1.0\times 10^6$ m/s. Snapshots are taken at 25ps, 50ps, 75ps and 100ps, showing that the skyrmion remains stable with time. }
\label{Fig6}
\end{figure}

Another important aspect of such injection of relativistic electrons from the Dirac Half-metal is the drift velocity of the skyrmion itself. The dynamics of a skyrmion can be effectively described by the Thiele equation, which balances the gyrotropic or Magnus force, damping force, and external driving forces given by~\cite{thiele1973steady},
\begin{equation}
\mathbf{G} \times (\mathbf{v}_{\text{Sk}} - \mathbf{v}_s) + D \mathbf{v}_{\text{Sk}} - \alpha D \mathbf{v}_{\text{Sk}} = 0
\end{equation}
 where \(\mathbf{G} = -4 \pi Q \frac{M_s}{\gamma} \hat{y}\) is the gyrotropic vector, \(Q = 1\) is the topological charge, \(M_s = 1.854 \times 10^{-23} \, \text{J/T}\) is the saturation magnetization, \(\gamma = 1.7608597 \times 10^{11} \, \text{rad/(s·T)}\) is the gyromagnetic ratio, \(\alpha = 0.01\) is the Gilbert damping parameter,  \(\mathbf{v}_{\text{Sk}}\) is the skyrmion velocity, and \(\mathbf{v}_s = 1 \times 10^5 \, \text{m/s}\) is the spin current velocity of incident electrons. We used a scalar value for the dissipative tensor, $D=D_0(1+\alpha)$, where $D_0 = 10^{-3} \, \text{kg/s}$,
 Considering our scenario where carriers are injected exclusively along the x-axis, we found that the skyrmion's velocity in the z direction is negligible, leading to a skyrmion Hall angle of almost equal to zero. This result implies that under the given conditions, the skyrmion moves strictly along the x-axis without any transverse motion, demonstrating the significant influence of damping forces in skyrmion dynamics within the xz-plane. This minimal transverse motion implies that the skyrmion moves predominantly along the direction of the applied current (for example, here along the x-direction), making this situation highly suitable for racetrack memory applications. In such devices, precise and directed skyrmion motion along the track is crucial to prevent data loss or misalignment, and the negligible Hall angle ensures that skyrmions remain confined to their intended paths, thus enhancing device reliability and performance.

In the end, we address some limitations of our study. Here we have considered the interaction between the relativistic dipole with the gauge field at the periphery of the skyrmion, where this effect is most pronounced. But it is also possible to investigate deep inside the skyrmion region where the matrix $\mathbf{R}$ is different from the unity matrix. It will be interesting to explore the torque behaviour at the \textit{very interior} of the skyrmion, which we keep for future work. Furthermore, the temperature effects are not considered as that would complicate the scenario. Lastly, the magnetic exchange interaction described by the Hamiltonian $H_{ex} = \int d \boldsymbol{r}^3 J(\nabla M)^2$ is omitted to avoid the complexities introduced by the parameter $J$, aiming to maintain the transparency of our results.

\section{Conclusions}
In this study, we have investigated the magnetization dynamics of a single skyrmion subjected to the injection of spin-polarized electrons originating from a Dirac half-metal (DHM). Dirac half-metals are characterized by their ability to emit electrons with drift velocities comparable to the Fermi velocity, leading to a substantial non-equilibrium magnetization density at the interface between the DHM and the skyrmion. The relativistic characteristics of these electrons mean that their magnetization density is associated with an electric dipole moment. This dipole moment interacts with the skyrmion's gauge electric field, resulting in an effective magnetic field. This field, in turn, exerts a torque on the skyrmion's magnetization texture.
To quantitatively understand these interactions, we modified the Landau-Lifshitz-Gilbert (LLG) equation to include the effects of spin torques driven by these high-velocity electron streams. Our modified LLG simulations reveal significant changes in the skyrmion's magnetic configurations over time. One of the most notable outcomes of our study is the observation that the skyrmion's topological charge can be substantially reduced by injecting electrons along a specific direction, effectively leading to the skyrmion's destruction. More importantly,  when relativistic spin-polarized carriers are injected orthogonally to the initial direction, the skyrmion's topological charge, and hence its characteristic magnetic texture is again restored.
These findings therefore highlight the potential for manipulating skyrmionic states using controlled electron injection. This capability suggests new avenues for employing skyrmion-based systems at the interfaces of topological materials and ferromagnets, potentially advancing applications in magnetic storage and spintronic devices.

\section*{Acknowledgments}
This work was supported by the Korea Institute of Science and Technology, GKP (Global Knowledge Platform, Grant number 2V6760) project of the Ministry of Science, ICT and Future Planning.

\section*{Data availability statement}
Any data that support the findings of this study are included within the article.
\bibliographystyle{apsrev4-1}
\bibliography{Ref.bib}

\begin{thebibliography}{47}%
\makeatletter
\providecommand \@ifxundefined [1]{%
 \@ifx{#1\undefined}
}%
\providecommand \@ifnum [1]{%
 \ifnum #1\expandafter \@firstoftwo
 \else \expandafter \@secondoftwo
 \fi
}%
\providecommand \@ifx [1]{%
 \ifx #1\expandafter \@firstoftwo
 \else \expandafter \@secondoftwo
 \fi
}%
\providecommand \natexlab [1]{#1}%
\providecommand \enquote  [1]{``#1''}%
\providecommand \bibnamefont  [1]{#1}%
\providecommand \bibfnamefont [1]{#1}%
\providecommand \citenamefont [1]{#1}%
\providecommand \href@noop [0]{\@secondoftwo}%
\providecommand \href [0]{\begingroup \@sanitize@url \@href}%
\providecommand \@href[1]{\@@startlink{#1}\@@href}%
\providecommand \@@href[1]{\endgroup#1\@@endlink}%
\providecommand \@sanitize@url [0]{\catcode `\\12\catcode `\$12\catcode
  `\&12\catcode `\#12\catcode `\^12\catcode `\_12\catcode `\%12\relax}%
\providecommand \@@startlink[1]{}%
\providecommand \@@endlink[0]{}%
\providecommand \url  [0]{\begingroup\@sanitize@url \@url }%
\providecommand \@url [1]{\endgroup\@href {#1}{\urlprefix }}%
\providecommand \urlprefix  [0]{URL }%
\providecommand \Eprint [0]{\href }%
\providecommand \doibase [0]{http://dx.doi.org/}%
\providecommand \selectlanguage [0]{\@gobble}%
\providecommand \bibinfo  [0]{\@secondoftwo}%
\providecommand \bibfield  [0]{\@secondoftwo}%
\providecommand \translation [1]{[#1]}%
\providecommand \BibitemOpen [0]{}%
\providecommand \bibitemStop [0]{}%
\providecommand \bibitemNoStop [0]{.\EOS\space}%
\providecommand \EOS [0]{\spacefactor3000\relax}%
\providecommand \BibitemShut  [1]{\csname bibitem#1\endcsname}%
\let\auto@bib@innerbib\@empty
\bibitem [{\citenamefont {Wang}\ \emph {et~al.}(2021)\citenamefont {Wang},
  \citenamefont {Hu},\ and\ \citenamefont {Wu}}]{text1}%
  \BibitemOpen
  \bibfield  {author} {\bibinfo {author} {\bibfnamefont {X.}~\bibnamefont
  {Wang}}, \bibinfo {author} {\bibfnamefont {X.~C.}\ \bibnamefont {Hu}}, \ and\
  \bibinfo {author} {\bibfnamefont {H.~T.}\ \bibnamefont {Wu}},\ }\href
  {\doibase 10.1038/S42005-021-00646-9} {\bibfield  {journal} {\bibinfo
  {journal} {Communications in Physics}\ }\textbf {\bibinfo {volume} {4}},\
  \bibinfo {pages} {142} (\bibinfo {year} {2021})}\BibitemShut {NoStop}%
\bibitem [{\citenamefont {Tveten}\ \emph {et~al.}(2016)\citenamefont {Tveten},
  \citenamefont {Muller}, \citenamefont {Linder},\ and\ \citenamefont
  {Brataas}}]{text2}%
  \BibitemOpen
  \bibfield  {author} {\bibinfo {author} {\bibfnamefont {E.~G.}\ \bibnamefont
  {Tveten}}, \bibinfo {author} {\bibfnamefont {T.}~\bibnamefont {Muller}},
  \bibinfo {author} {\bibfnamefont {J.}~\bibnamefont {Linder}}, \ and\ \bibinfo
  {author} {\bibfnamefont {A.}~\bibnamefont {Brataas}},\ }\href {\doibase
  10.1103/PHYSREVB.93.104408} {\bibfield  {journal} {\bibinfo  {journal}
  {Physical Review B}\ }\textbf {\bibinfo {volume} {93}},\ \bibinfo {pages}
  {104408} (\bibinfo {year} {2016})}\BibitemShut {NoStop}%
\bibitem [{\citenamefont {Antos}\ \emph {et~al.}(2008)\citenamefont {Antos},
  \citenamefont {Otani},\ and\ \citenamefont {Shibata}}]{antos2008magnetic}%
  \BibitemOpen
  \bibfield  {author} {\bibinfo {author} {\bibfnamefont {R.}~\bibnamefont
  {Antos}}, \bibinfo {author} {\bibfnamefont {Y.}~\bibnamefont {Otani}}, \ and\
  \bibinfo {author} {\bibfnamefont {J.}~\bibnamefont {Shibata}},\ }\href@noop
  {} {\bibfield  {journal} {\bibinfo  {journal} {Journal of the Physical
  Society of Japan}\ }\textbf {\bibinfo {volume} {77}},\ \bibinfo {pages}
  {031004} (\bibinfo {year} {2008})}\BibitemShut {NoStop}%
\bibitem [{\citenamefont {Fert}\ \emph {et~al.}(2017)\citenamefont {Fert},
  \citenamefont {Reyren},\ and\ \citenamefont {Cros}}]{sk1}%
  \BibitemOpen
  \bibfield  {author} {\bibinfo {author} {\bibfnamefont {A.}~\bibnamefont
  {Fert}}, \bibinfo {author} {\bibfnamefont {N.}~\bibnamefont {Reyren}}, \ and\
  \bibinfo {author} {\bibfnamefont {V.}~\bibnamefont {Cros}},\ }\href {\doibase
  10.1038/NATREVMATS.2017.31} {\bibfield  {journal} {\bibinfo  {journal}
  {Nature Reviews Materials}\ }\textbf {\bibinfo {volume} {2}},\ \bibinfo
  {pages} {17031} (\bibinfo {year} {2017})}\BibitemShut {NoStop}%
\bibitem [{\citenamefont {Wiesendanger}(2016)}]{sk2}%
  \BibitemOpen
  \bibfield  {author} {\bibinfo {author} {\bibfnamefont {R.}~\bibnamefont
  {Wiesendanger}},\ }\href {\doibase 10.1038/NATREVMATS.2016.44} {\bibfield
  {journal} {\bibinfo  {journal} {Nature Reviews Materials}\ }\textbf {\bibinfo
  {volume} {1}},\ \bibinfo {pages} {16044} (\bibinfo {year}
  {2016})}\BibitemShut {NoStop}%
\bibitem [{\citenamefont {Nagaosa}\ \emph {et~al.}(2013)\citenamefont
  {Nagaosa}, \citenamefont {Tokura},\ and\ \citenamefont {Tokura}}]{sk3}%
  \BibitemOpen
  \bibfield  {author} {\bibinfo {author} {\bibfnamefont {N.}~\bibnamefont
  {Nagaosa}}, \bibinfo {author} {\bibfnamefont {Y.}~\bibnamefont {Tokura}}, \
  and\ \bibinfo {author} {\bibfnamefont {Y.}~\bibnamefont {Tokura}},\ }\href
  {\doibase 10.1038/NNANO.2013.243} {\bibfield  {journal} {\bibinfo  {journal}
  {Nature Nanotechnology}\ }\textbf {\bibinfo {volume} {8}},\ \bibinfo {pages}
  {899} (\bibinfo {year} {2013})}\BibitemShut {NoStop}%
\bibitem [{\citenamefont {Fert}\ \emph {et~al.}(2013)\citenamefont {Fert},
  \citenamefont {Cros},\ and\ \citenamefont {Sampaio}}]{Fert2013SkyrmionsOT}%
  \BibitemOpen
  \bibfield  {author} {\bibinfo {author} {\bibfnamefont {A.}~\bibnamefont
  {Fert}}, \bibinfo {author} {\bibfnamefont {V.}~\bibnamefont {Cros}}, \ and\
  \bibinfo {author} {\bibfnamefont {J.}~\bibnamefont {Sampaio}},\ }\href
  {https://api.semanticscholar.org/CorpusID:42822349} {\bibfield  {journal}
  {\bibinfo  {journal} {Nature nanotechnology}\ }\textbf {\bibinfo {volume} {8
  3}},\ \bibinfo {pages} {152} (\bibinfo {year} {2013})}\BibitemShut {NoStop}%
\bibitem [{\citenamefont {Zhang}\ \emph {et~al.}(2023)\citenamefont {Zhang},
  \citenamefont {Zhang}, \citenamefont {Hou}, \citenamefont {Qin},
  \citenamefont {Gao},\ and\ \citenamefont {Liu}}]{zhang2023}%
  \BibitemOpen
  \bibfield  {author} {\bibinfo {author} {\bibfnamefont {H.}~\bibnamefont
  {Zhang}}, \bibinfo {author} {\bibfnamefont {Y.}~\bibnamefont {Zhang}},
  \bibinfo {author} {\bibfnamefont {Z.}~\bibnamefont {Hou}}, \bibinfo {author}
  {\bibfnamefont {M.}~\bibnamefont {Qin}}, \bibinfo {author} {\bibfnamefont
  {X.}~\bibnamefont {Gao}}, \ and\ \bibinfo {author} {\bibfnamefont
  {J.}~\bibnamefont {Liu}},\ }\href@noop {} {\bibfield  {journal} {\bibinfo
  {journal} {Materials Futures}\ }\textbf {\bibinfo {volume} {2}},\ \bibinfo
  {pages} {032201} (\bibinfo {year} {2023})}\BibitemShut {NoStop}%
\bibitem [{\citenamefont {Yang}\ \emph {et~al.}(2024)\citenamefont {Yang},
  \citenamefont {Zhao}, \citenamefont {Yi}, \citenamefont {Xu}, \citenamefont
  {Chai}, \citenamefont {Zhang}, \citenamefont {Jiang}, \citenamefont {Ji},
  \citenamefont {Hou}, \citenamefont {Jiang} \emph
  {et~al.}}]{yang2024acoustic}%
  \BibitemOpen
  \bibfield  {author} {\bibinfo {author} {\bibfnamefont {Y.}~\bibnamefont
  {Yang}}, \bibinfo {author} {\bibfnamefont {L.}~\bibnamefont {Zhao}}, \bibinfo
  {author} {\bibfnamefont {D.}~\bibnamefont {Yi}}, \bibinfo {author}
  {\bibfnamefont {T.}~\bibnamefont {Xu}}, \bibinfo {author} {\bibfnamefont
  {Y.}~\bibnamefont {Chai}}, \bibinfo {author} {\bibfnamefont {C.}~\bibnamefont
  {Zhang}}, \bibinfo {author} {\bibfnamefont {D.}~\bibnamefont {Jiang}},
  \bibinfo {author} {\bibfnamefont {Y.}~\bibnamefont {Ji}}, \bibinfo {author}
  {\bibfnamefont {D.}~\bibnamefont {Hou}}, \bibinfo {author} {\bibfnamefont
  {W.}~\bibnamefont {Jiang}},  \emph {et~al.},\ }\href@noop {} {\bibfield
  {journal} {\bibinfo  {journal} {Nature Communications}\ }\textbf {\bibinfo
  {volume} {15}},\ \bibinfo {pages} {1018} (\bibinfo {year}
  {2024})}\BibitemShut {NoStop}%
\bibitem [{\citenamefont {Hu}\ \emph {et~al.}(2017)\citenamefont {Hu},
  \citenamefont {Chi}, \citenamefont {Li}, \citenamefont {Liu},\ and\
  \citenamefont {Du}}]{hu2017creation}%
  \BibitemOpen
  \bibfield  {author} {\bibinfo {author} {\bibfnamefont {Y.}~\bibnamefont
  {Hu}}, \bibinfo {author} {\bibfnamefont {X.}~\bibnamefont {Chi}}, \bibinfo
  {author} {\bibfnamefont {X.}~\bibnamefont {Li}}, \bibinfo {author}
  {\bibfnamefont {Y.}~\bibnamefont {Liu}}, \ and\ \bibinfo {author}
  {\bibfnamefont {A.}~\bibnamefont {Du}},\ }\href@noop {} {\bibfield  {journal}
  {\bibinfo  {journal} {Scientific Reports}\ }\textbf {\bibinfo {volume} {7}},\
  \bibinfo {pages} {16079} (\bibinfo {year} {2017})}\BibitemShut {NoStop}%
\bibitem [{\citenamefont {Koshibae}\ and\ \citenamefont
  {Nagaosa}(2014)}]{koshibae2014creation}%
  \BibitemOpen
  \bibfield  {author} {\bibinfo {author} {\bibfnamefont {W.}~\bibnamefont
  {Koshibae}}\ and\ \bibinfo {author} {\bibfnamefont {N.}~\bibnamefont
  {Nagaosa}},\ }\href@noop {} {\bibfield  {journal} {\bibinfo  {journal}
  {Nature communications}\ }\textbf {\bibinfo {volume} {5}},\ \bibinfo {pages}
  {5148} (\bibinfo {year} {2014})}\BibitemShut {NoStop}%
\bibitem [{\citenamefont {Pesin}\ and\ \citenamefont
  {MacDonald}(2012)}]{pesin2012spintronics}%
  \BibitemOpen
  \bibfield  {author} {\bibinfo {author} {\bibfnamefont {D.}~\bibnamefont
  {Pesin}}\ and\ \bibinfo {author} {\bibfnamefont {A.~H.}\ \bibnamefont
  {MacDonald}},\ }\href@noop {} {\bibfield  {journal} {\bibinfo  {journal}
  {Nature materials}\ }\textbf {\bibinfo {volume} {11}},\ \bibinfo {pages}
  {409} (\bibinfo {year} {2012})}\BibitemShut {NoStop}%
\bibitem [{\citenamefont {Baker}\ \emph {et~al.}(2015)\citenamefont {Baker},
  \citenamefont {Figueroa}, \citenamefont {Collins-McIntyre}, \citenamefont
  {Van Der~Laan},\ and\ \citenamefont {Hesjedal}}]{baker2015spin}%
  \BibitemOpen
  \bibfield  {author} {\bibinfo {author} {\bibfnamefont {A.}~\bibnamefont
  {Baker}}, \bibinfo {author} {\bibfnamefont {A.}~\bibnamefont {Figueroa}},
  \bibinfo {author} {\bibfnamefont {L.}~\bibnamefont {Collins-McIntyre}},
  \bibinfo {author} {\bibfnamefont {G.}~\bibnamefont {Van Der~Laan}}, \ and\
  \bibinfo {author} {\bibfnamefont {T.}~\bibnamefont {Hesjedal}},\ }\href@noop
  {} {\bibfield  {journal} {\bibinfo  {journal} {Scientific reports}\ }\textbf
  {\bibinfo {volume} {5}},\ \bibinfo {pages} {7907} (\bibinfo {year}
  {2015})}\BibitemShut {NoStop}%
\bibitem [{\citenamefont {Jonietz}\ \emph {et~al.}(2010)\citenamefont
  {Jonietz}, \citenamefont {M{\"u}hlbauer}, \citenamefont {Pfleiderer},
  \citenamefont {Neubauer}, \citenamefont {M{\"u}nzer}, \citenamefont {Bauer},
  \citenamefont {Adams}, \citenamefont {Georgii}, \citenamefont {B{\"o}ni},
  \citenamefont {Duine} \emph {et~al.}}]{jonietz2010spin}%
  \BibitemOpen
  \bibfield  {author} {\bibinfo {author} {\bibfnamefont {F.}~\bibnamefont
  {Jonietz}}, \bibinfo {author} {\bibfnamefont {S.}~\bibnamefont
  {M{\"u}hlbauer}}, \bibinfo {author} {\bibfnamefont {C.}~\bibnamefont
  {Pfleiderer}}, \bibinfo {author} {\bibfnamefont {A.}~\bibnamefont
  {Neubauer}}, \bibinfo {author} {\bibfnamefont {W.}~\bibnamefont
  {M{\"u}nzer}}, \bibinfo {author} {\bibfnamefont {A.}~\bibnamefont {Bauer}},
  \bibinfo {author} {\bibfnamefont {T.}~\bibnamefont {Adams}}, \bibinfo
  {author} {\bibfnamefont {R.}~\bibnamefont {Georgii}}, \bibinfo {author}
  {\bibfnamefont {P.}~\bibnamefont {B{\"o}ni}}, \bibinfo {author}
  {\bibfnamefont {R.~A.}\ \bibnamefont {Duine}},  \emph {et~al.},\ }\href@noop
  {} {\bibfield  {journal} {\bibinfo  {journal} {Science}\ }\textbf {\bibinfo
  {volume} {330}},\ \bibinfo {pages} {1648} (\bibinfo {year}
  {2010})}\BibitemShut {NoStop}%
\bibitem [{\citenamefont {Manchon}\ \emph {et~al.}(2015)\citenamefont
  {Manchon}, \citenamefont {Koo}, \citenamefont {Nitta}, \citenamefont
  {Frolov},\ and\ \citenamefont {Duine}}]{manchon2015new}%
  \BibitemOpen
  \bibfield  {author} {\bibinfo {author} {\bibfnamefont {A.}~\bibnamefont
  {Manchon}}, \bibinfo {author} {\bibfnamefont {H.~C.}\ \bibnamefont {Koo}},
  \bibinfo {author} {\bibfnamefont {J.}~\bibnamefont {Nitta}}, \bibinfo
  {author} {\bibfnamefont {S.~M.}\ \bibnamefont {Frolov}}, \ and\ \bibinfo
  {author} {\bibfnamefont {R.~A.}\ \bibnamefont {Duine}},\ }\href@noop {}
  {\bibfield  {journal} {\bibinfo  {journal} {Nature materials}\ }\textbf
  {\bibinfo {volume} {14}},\ \bibinfo {pages} {871} (\bibinfo {year}
  {2015})}\BibitemShut {NoStop}%
\bibitem [{\citenamefont {Chen}\ \emph {et~al.}(2019)\citenamefont {Chen},
  \citenamefont {Wang}, \citenamefont {Zhang}, \citenamefont {Zhou},
  \citenamefont {Zhang}, \citenamefont {Jin}, \citenamefont {sen Wang},
  \citenamefont {Qin}, \citenamefont {Qiu}, \citenamefont {Mei}, \citenamefont
  {Ye}, \citenamefont {Xi}, \citenamefont {He}, \citenamefont {Li},\ and\
  \citenamefont {Wang}}]{Chen2019Evidence}%
  \BibitemOpen
  \bibfield  {author} {\bibinfo {author} {\bibfnamefont {J.}~\bibnamefont
  {Chen}}, \bibinfo {author} {\bibfnamefont {L.}~\bibnamefont {Wang}}, \bibinfo
  {author} {\bibfnamefont {M.}~\bibnamefont {Zhang}}, \bibinfo {author}
  {\bibfnamefont {L.}~\bibnamefont {Zhou}}, \bibinfo {author} {\bibfnamefont
  {R.}~\bibnamefont {Zhang}}, \bibinfo {author} {\bibfnamefont
  {L.}~\bibnamefont {Jin}}, \bibinfo {author} {\bibfnamefont {X.}~\bibnamefont
  {sen Wang}}, \bibinfo {author} {\bibfnamefont {H.}~\bibnamefont {Qin}},
  \bibinfo {author} {\bibfnamefont {Y.}~\bibnamefont {Qiu}}, \bibinfo {author}
  {\bibfnamefont {J.}~\bibnamefont {Mei}}, \bibinfo {author} {\bibfnamefont
  {F.}~\bibnamefont {Ye}}, \bibinfo {author} {\bibfnamefont {B.}~\bibnamefont
  {Xi}}, \bibinfo {author} {\bibfnamefont {H.}~\bibnamefont {He}}, \bibinfo
  {author} {\bibfnamefont {B.}~\bibnamefont {Li}}, \ and\ \bibinfo {author}
  {\bibfnamefont {G.}~\bibnamefont {Wang}},\ }\href {\doibase
  10.1021/acs.nanolett.9b02191} {\bibfield  {journal} {\bibinfo  {journal}
  {Nano letters}\ } (\bibinfo {year} {2019}),\
  10.1021/acs.nanolett.9b02191}\BibitemShut {NoStop}%
\bibitem [{\citenamefont {Dahir}\ \emph {et~al.}(2018)\citenamefont {Dahir},
  \citenamefont {Volkov},\ and\ \citenamefont {Eremin}}]{Dahir2018Interaction}%
  \BibitemOpen
  \bibfield  {author} {\bibinfo {author} {\bibfnamefont {S.~M.}\ \bibnamefont
  {Dahir}}, \bibinfo {author} {\bibfnamefont {A.}~\bibnamefont {Volkov}}, \
  and\ \bibinfo {author} {\bibfnamefont {I.}~\bibnamefont {Eremin}},\ }\href
  {\doibase 10.1103/PhysRevLett.122.097001} {\bibfield  {journal} {\bibinfo
  {journal} {Physical review letters}\ }\textbf {\bibinfo {volume} {122 9}},\
  \bibinfo {pages} {097001} (\bibinfo {year} {2018})}\BibitemShut {NoStop}%
\bibitem [{\citenamefont {Zarzuela}\ \emph {et~al.}(2019)\citenamefont
  {Zarzuela}, \citenamefont {Bharadwaj}, \citenamefont {Kim}, \citenamefont
  {Sinova},\ and\ \citenamefont {Everschor-Sitte}}]{Zarzuela2019Stability}%
  \BibitemOpen
  \bibfield  {author} {\bibinfo {author} {\bibfnamefont {R.}~\bibnamefont
  {Zarzuela}}, \bibinfo {author} {\bibfnamefont {V.~K.}\ \bibnamefont
  {Bharadwaj}}, \bibinfo {author} {\bibfnamefont {K.-W.}\ \bibnamefont {Kim}},
  \bibinfo {author} {\bibfnamefont {J.}~\bibnamefont {Sinova}}, \ and\ \bibinfo
  {author} {\bibfnamefont {K.}~\bibnamefont {Everschor-Sitte}},\ }\href
  {\doibase 10.1103/PhysRevB.101.054405} {\bibfield  {journal} {\bibinfo
  {journal} {Physical Review B}\ } (\bibinfo {year} {2019}),\
  10.1103/PhysRevB.101.054405}\BibitemShut {NoStop}%
\bibitem [{\citenamefont {Dou}\ \emph {et~al.}(2023)\citenamefont {Dou},
  \citenamefont {Zhang}, \citenamefont {Guo}, \citenamefont {Zhu},
  \citenamefont {Luo}, \citenamefont {Zhao}, \citenamefont {Huang},
  \citenamefont {Yu}, \citenamefont {Zhao}, \citenamefont {Qi}, \citenamefont
  {Deng}, \citenamefont {Wang}, \citenamefont {Li}, \citenamefont {Shen},
  \citenamefont {Zheng}, \citenamefont {Wu}, \citenamefont {Yang},
  \citenamefont {ben Shen},\ and\ \citenamefont {Wang}}]{Dou2023Deterministic}%
  \BibitemOpen
  \bibfield  {author} {\bibinfo {author} {\bibfnamefont {P.}~\bibnamefont
  {Dou}}, \bibinfo {author} {\bibfnamefont {J.}~\bibnamefont {Zhang}}, \bibinfo
  {author} {\bibfnamefont {Y.}~\bibnamefont {Guo}}, \bibinfo {author}
  {\bibfnamefont {T.}~\bibnamefont {Zhu}}, \bibinfo {author} {\bibfnamefont
  {J.}~\bibnamefont {Luo}}, \bibinfo {author} {\bibfnamefont {G.}~\bibnamefont
  {Zhao}}, \bibinfo {author} {\bibfnamefont {H.}~\bibnamefont {Huang}},
  \bibinfo {author} {\bibfnamefont {G.}~\bibnamefont {Yu}}, \bibinfo {author}
  {\bibfnamefont {Y.}~\bibnamefont {Zhao}}, \bibinfo {author} {\bibfnamefont
  {J.}~\bibnamefont {Qi}}, \bibinfo {author} {\bibfnamefont {X.}~\bibnamefont
  {Deng}}, \bibinfo {author} {\bibfnamefont {Y.}~\bibnamefont {Wang}}, \bibinfo
  {author} {\bibfnamefont {J.}~\bibnamefont {Li}}, \bibinfo {author}
  {\bibfnamefont {J.}~\bibnamefont {Shen}}, \bibinfo {author} {\bibfnamefont
  {X.}~\bibnamefont {Zheng}}, \bibinfo {author} {\bibfnamefont
  {Y.}~\bibnamefont {Wu}}, \bibinfo {author} {\bibfnamefont {H.}~\bibnamefont
  {Yang}}, \bibinfo {author} {\bibfnamefont {B.}~\bibnamefont {ben Shen}}, \
  and\ \bibinfo {author} {\bibfnamefont {S.}~\bibnamefont {Wang}},\ }\href
  {\doibase 10.1021/acs.nanolett.3c01192} {\bibfield  {journal} {\bibinfo
  {journal} {Nano letters}\ } (\bibinfo {year} {2023}),\
  10.1021/acs.nanolett.3c01192}\BibitemShut {NoStop}%
\bibitem [{\citenamefont {Saini}\ \emph {et~al.}(2023)\citenamefont {Saini},
  \citenamefont {Shukla}, \citenamefont {Bindal},\ and\ \citenamefont
  {Kaushik}}]{Saini2023Spin}%
  \BibitemOpen
  \bibfield  {author} {\bibinfo {author} {\bibfnamefont {S.}~\bibnamefont
  {Saini}}, \bibinfo {author} {\bibfnamefont {A.}~\bibnamefont {Shukla}},
  \bibinfo {author} {\bibfnamefont {N.}~\bibnamefont {Bindal}}, \ and\ \bibinfo
  {author} {\bibfnamefont {B.}~\bibnamefont {Kaushik}},\ }\href {\doibase
  10.1109/NMDC57951.2023.10344136} {\bibfield  {journal} {\bibinfo  {journal}
  {2023 IEEE Nanotechnology Materials and Devices Conference (NMDC)}\ ,\
  \bibinfo {pages} {47}} (\bibinfo {year} {2023})}\BibitemShut {NoStop}%
\bibitem [{\citenamefont {Li}\ \emph {et~al.}(2019)\citenamefont {Li},
  \citenamefont {Dong}, \citenamefont {Han}, \citenamefont {Song},
  \citenamefont {Li}, \citenamefont {Zhu}, \citenamefont {Li},\ and\
  \citenamefont {Sun}}]{Li2019Micromagnetic}%
  \BibitemOpen
  \bibfield  {author} {\bibinfo {author} {\bibfnamefont {L.}~\bibnamefont
  {Li}}, \bibinfo {author} {\bibfnamefont {S.}~\bibnamefont {Dong}}, \bibinfo
  {author} {\bibfnamefont {R.}~\bibnamefont {Han}}, \bibinfo {author}
  {\bibfnamefont {K.}~\bibnamefont {Song}}, \bibinfo {author} {\bibfnamefont
  {D.}~\bibnamefont {Li}}, \bibinfo {author} {\bibfnamefont {M.}~\bibnamefont
  {Zhu}}, \bibinfo {author} {\bibfnamefont {W.}~\bibnamefont {Li}}, \ and\
  \bibinfo {author} {\bibfnamefont {W.}~\bibnamefont {Sun}},\ }\href@noop {}
  {\bibfield  {journal} {\bibinfo  {journal} {Journal of Rare Earths}\ }\textbf
  {\bibinfo {volume} {37}},\ \bibinfo {pages} {628} (\bibinfo {year}
  {2019})}\BibitemShut {NoStop}%
\bibitem [{\citenamefont {Gilbert}(2004)}]{gilbert2004phenomenological}%
  \BibitemOpen
  \bibfield  {author} {\bibinfo {author} {\bibfnamefont {T.~L.}\ \bibnamefont
  {Gilbert}},\ }\href@noop {} {\bibfield  {journal} {\bibinfo  {journal} {IEEE
  transactions on magnetics}\ }\textbf {\bibinfo {volume} {40}},\ \bibinfo
  {pages} {3443} (\bibinfo {year} {2004})}\BibitemShut {NoStop}%
\bibitem [{\citenamefont {Mahfouzi}\ \emph {et~al.}(2016)\citenamefont
  {Mahfouzi}, \citenamefont {Nikoli{\'c}},\ and\ \citenamefont
  {Kioussis}}]{mahfouzi2016antidamping}%
  \BibitemOpen
  \bibfield  {author} {\bibinfo {author} {\bibfnamefont {F.}~\bibnamefont
  {Mahfouzi}}, \bibinfo {author} {\bibfnamefont {B.~K.}\ \bibnamefont
  {Nikoli{\'c}}}, \ and\ \bibinfo {author} {\bibfnamefont {N.}~\bibnamefont
  {Kioussis}},\ }\href@noop {} {\bibfield  {journal} {\bibinfo  {journal}
  {Physical Review B}\ }\textbf {\bibinfo {volume} {93}},\ \bibinfo {pages}
  {115419} (\bibinfo {year} {2016})}\BibitemShut {NoStop}%
\bibitem [{\citenamefont {Haidar}\ \emph {et~al.}(2019)\citenamefont {Haidar},
  \citenamefont {Awad}, \citenamefont {Dvornik}, \citenamefont {Khymyn},
  \citenamefont {Houshang},\ and\ \citenamefont
  {{\AA}kerman}}]{haidar2019single}%
  \BibitemOpen
  \bibfield  {author} {\bibinfo {author} {\bibfnamefont {M.}~\bibnamefont
  {Haidar}}, \bibinfo {author} {\bibfnamefont {A.~A.}\ \bibnamefont {Awad}},
  \bibinfo {author} {\bibfnamefont {M.}~\bibnamefont {Dvornik}}, \bibinfo
  {author} {\bibfnamefont {R.}~\bibnamefont {Khymyn}}, \bibinfo {author}
  {\bibfnamefont {A.}~\bibnamefont {Houshang}}, \ and\ \bibinfo {author}
  {\bibfnamefont {J.}~\bibnamefont {{\AA}kerman}},\ }\href@noop {} {\bibfield
  {journal} {\bibinfo  {journal} {Nature communications}\ }\textbf {\bibinfo
  {volume} {10}},\ \bibinfo {pages} {2362} (\bibinfo {year}
  {2019})}\BibitemShut {NoStop}%
\bibitem [{\citenamefont {Bhattacharjee}\ and\ \citenamefont
  {Lee}(2019)}]{bhattacharjee2019first}%
  \BibitemOpen
  \bibfield  {author} {\bibinfo {author} {\bibfnamefont {S.}~\bibnamefont
  {Bhattacharjee}}\ and\ \bibinfo {author} {\bibfnamefont {S.-C.}\ \bibnamefont
  {Lee}},\ }\href@noop {} {\bibfield  {journal} {\bibinfo  {journal}
  {Scientific Reports}\ }\textbf {\bibinfo {volume} {9}},\ \bibinfo {pages}
  {8381} (\bibinfo {year} {2019})}\BibitemShut {NoStop}%
\bibitem [{\citenamefont {Chen}\ \emph {et~al.}(2022)\citenamefont {Chen},
  \citenamefont {Ni}, \citenamefont {Yao},\ and\ \citenamefont
  {Hou}}]{chen2022}%
  \BibitemOpen
  \bibfield  {author} {\bibinfo {author} {\bibfnamefont {C.-Q.}\ \bibnamefont
  {Chen}}, \bibinfo {author} {\bibfnamefont {X.-S.}\ \bibnamefont {Ni}},
  \bibinfo {author} {\bibfnamefont {D.-X.}\ \bibnamefont {Yao}}, \ and\
  \bibinfo {author} {\bibfnamefont {Y.}~\bibnamefont {Hou}},\ }\href@noop {}
  {\bibfield  {journal} {\bibinfo  {journal} {Applied Physics Letters}\
  }\textbf {\bibinfo {volume} {121}} (\bibinfo {year} {2022})}\BibitemShut
  {NoStop}%
\bibitem [{\citenamefont {Wang}\ \emph {et~al.}(2018)\citenamefont {Wang},
  \citenamefont {Li}, \citenamefont {Zhang}, \citenamefont {Zhang},
  \citenamefont {Ji}, \citenamefont {Li},\ and\ \citenamefont
  {Wang}}]{wang2018high}%
  \BibitemOpen
  \bibfield  {author} {\bibinfo {author} {\bibfnamefont {Y.}~\bibnamefont
  {Wang}}, \bibinfo {author} {\bibfnamefont {S.}~\bibnamefont {Li}}, \bibinfo
  {author} {\bibfnamefont {C.}~\bibnamefont {Zhang}}, \bibinfo {author}
  {\bibfnamefont {S.}~\bibnamefont {Zhang}}, \bibinfo {author} {\bibfnamefont
  {W.}~\bibnamefont {Ji}}, \bibinfo {author} {\bibfnamefont {P.}~\bibnamefont
  {Li}}, \ and\ \bibinfo {author} {\bibfnamefont {P.}~\bibnamefont {Wang}},\
  }\href {\doibase 10.1039/C8TC02500B} {\bibfield  {journal} {\bibinfo
  {journal} {Journal of Materials Chemistry C}\ }\textbf {\bibinfo {volume}
  {6}},\ \bibinfo {pages} {10284} (\bibinfo {year} {2018})}\BibitemShut
  {NoStop}%
\bibitem [{\citenamefont {Bhattacharjee}\ and\ \citenamefont
  {Lee}(2023)}]{bhattacharjee2023spin}%
  \BibitemOpen
  \bibfield  {author} {\bibinfo {author} {\bibfnamefont {S.}~\bibnamefont
  {Bhattacharjee}}\ and\ \bibinfo {author} {\bibfnamefont {S.-C.}\ \bibnamefont
  {Lee}},\ }\href@noop {} {\bibfield  {journal} {\bibinfo  {journal} {Journal
  of Physics: Condensed Matter}\ }\textbf {\bibinfo {volume} {35}},\ \bibinfo
  {pages} {435802} (\bibinfo {year} {2023})}\BibitemShut {NoStop}%
\bibitem [{\citenamefont {B{\"u}ttner}\ \emph {et~al.}(2018)\citenamefont
  {B{\"u}ttner}, \citenamefont {Lemesh},\ and\ \citenamefont
  {Beach}}]{buttner2018theory}%
  \BibitemOpen
  \bibfield  {author} {\bibinfo {author} {\bibfnamefont {F.}~\bibnamefont
  {B{\"u}ttner}}, \bibinfo {author} {\bibfnamefont {I.}~\bibnamefont {Lemesh}},
  \ and\ \bibinfo {author} {\bibfnamefont {G.~S.}\ \bibnamefont {Beach}},\
  }\href@noop {} {\bibfield  {journal} {\bibinfo  {journal} {Scientific
  reports}\ }\textbf {\bibinfo {volume} {8}},\ \bibinfo {pages} {4464}
  (\bibinfo {year} {2018})}\BibitemShut {NoStop}%
\bibitem [{\citenamefont {Liu}\ \emph {et~al.}(2017)\citenamefont {Liu},
  \citenamefont {Liu},\ and\ \citenamefont {Zhao}}]{liu2017yn}%
  \BibitemOpen
  \bibfield  {author} {\bibinfo {author} {\bibfnamefont {Z.}~\bibnamefont
  {Liu}}, \bibinfo {author} {\bibfnamefont {J.}~\bibnamefont {Liu}}, \ and\
  \bibinfo {author} {\bibfnamefont {J.}~\bibnamefont {Zhao}},\ }\href@noop {}
  {\bibfield  {journal} {\bibinfo  {journal} {Nano Research}\ }\textbf
  {\bibinfo {volume} {10}},\ \bibinfo {pages} {1972} (\bibinfo {year}
  {2017})}\BibitemShut {NoStop}%
\bibitem [{\citenamefont {Ma}\ \emph {et~al.}(2018)\citenamefont {Ma},
  \citenamefont {Jiao}, \citenamefont {Jiang},\ and\ \citenamefont
  {Du}}]{ma2018rhombohedral}%
  \BibitemOpen
  \bibfield  {author} {\bibinfo {author} {\bibfnamefont {F.}~\bibnamefont
  {Ma}}, \bibinfo {author} {\bibfnamefont {Y.}~\bibnamefont {Jiao}}, \bibinfo
  {author} {\bibfnamefont {Z.}~\bibnamefont {Jiang}}, \ and\ \bibinfo {author}
  {\bibfnamefont {A.}~\bibnamefont {Du}},\ }\href@noop {} {\bibfield  {journal}
  {\bibinfo  {journal} {ACS applied materials \& interfaces}\ }\textbf
  {\bibinfo {volume} {10}},\ \bibinfo {pages} {36088} (\bibinfo {year}
  {2018})}\BibitemShut {NoStop}%
\bibitem [{\citenamefont {Ishizuka}\ and\ \citenamefont
  {Motome}(2012)}]{ishizuka2012dirac}%
  \BibitemOpen
  \bibfield  {author} {\bibinfo {author} {\bibfnamefont {H.}~\bibnamefont
  {Ishizuka}}\ and\ \bibinfo {author} {\bibfnamefont {Y.}~\bibnamefont
  {Motome}},\ }\href@noop {} {\bibfield  {journal} {\bibinfo  {journal}
  {Physical Review Letters}\ }\textbf {\bibinfo {volume} {109}},\ \bibinfo
  {pages} {237207} (\bibinfo {year} {2012})}\BibitemShut {NoStop}%
\bibitem [{\citenamefont {Panofsky}\ and\ \citenamefont
  {Phillips}(2005)}]{panofsky2005classical}%
  \BibitemOpen
  \bibfield  {author} {\bibinfo {author} {\bibfnamefont {W.~K.}\ \bibnamefont
  {Panofsky}}\ and\ \bibinfo {author} {\bibfnamefont {M.}~\bibnamefont
  {Phillips}},\ }\href@noop {} {\emph {\bibinfo {title} {Classical electricity
  and magnetism}}}\ (\bibinfo  {publisher} {Courier Corporation},\ \bibinfo
  {year} {2005})\BibitemShut {NoStop}%
\bibitem [{\citenamefont {Vekstein}(2011)}]{vekstein2011comment}%
  \BibitemOpen
  \bibfield  {author} {\bibinfo {author} {\bibfnamefont {G.}~\bibnamefont
  {Vekstein}},\ }\href@noop {} {\bibfield  {journal} {\bibinfo  {journal}
  {European Journal of Physics}\ }\textbf {\bibinfo {volume} {33}},\ \bibinfo
  {pages} {L1} (\bibinfo {year} {2011})}\BibitemShut {NoStop}%
\bibitem [{\citenamefont {Griffiths}\ and\ \citenamefont
  {Hnizdo}(2013)}]{griffiths2013mansuripur}%
  \BibitemOpen
  \bibfield  {author} {\bibinfo {author} {\bibfnamefont {D.~J.}\ \bibnamefont
  {Griffiths}}\ and\ \bibinfo {author} {\bibfnamefont {V.}~\bibnamefont
  {Hnizdo}},\ }\href@noop {} {\bibfield  {journal} {\bibinfo  {journal}
  {American Journal of Physics}\ }\textbf {\bibinfo {volume} {81}},\ \bibinfo
  {pages} {570} (\bibinfo {year} {2013})}\BibitemShut {NoStop}%
\bibitem [{\citenamefont {Matos-Abiague}\ and\ \citenamefont
  {Rodriguez-Suarez}(2009)}]{matos2009spin}%
  \BibitemOpen
  \bibfield  {author} {\bibinfo {author} {\bibfnamefont {A.}~\bibnamefont
  {Matos-Abiague}}\ and\ \bibinfo {author} {\bibfnamefont {R.}~\bibnamefont
  {Rodriguez-Suarez}},\ }\href@noop {} {\bibfield  {journal} {\bibinfo
  {journal} {Physical Review B}\ }\textbf {\bibinfo {volume} {80}},\ \bibinfo
  {pages} {094424} (\bibinfo {year} {2009})}\BibitemShut {NoStop}%
\bibitem [{\citenamefont {Tatara}(2019)}]{tatara2019effective}%
  \BibitemOpen
  \bibfield  {author} {\bibinfo {author} {\bibfnamefont {G.}~\bibnamefont
  {Tatara}},\ }\href@noop {} {\bibfield  {journal} {\bibinfo  {journal}
  {Physica E: Low-dimensional Systems and Nanostructures}\ }\textbf {\bibinfo
  {volume} {106}},\ \bibinfo {pages} {208} (\bibinfo {year}
  {2019})}\BibitemShut {NoStop}%
\bibitem [{\citenamefont {Araki}(2020)}]{araki2020magnetic}%
  \BibitemOpen
  \bibfield  {author} {\bibinfo {author} {\bibfnamefont {Y.}~\bibnamefont
  {Araki}},\ }\href@noop {} {\bibfield  {journal} {\bibinfo  {journal} {Annalen
  der Physik}\ }\textbf {\bibinfo {volume} {532}},\ \bibinfo {pages} {1900287}
  (\bibinfo {year} {2020})}\BibitemShut {NoStop}%
\bibitem [{\citenamefont {Jalil}\ \emph
  {et~al.}(2014{\natexlab{a}})\citenamefont {Jalil}, \citenamefont {Ghee~Tan},
  \citenamefont {Eason},\ and\ \citenamefont {Kong}}]{jalil2014topological}%
  \BibitemOpen
  \bibfield  {author} {\bibinfo {author} {\bibfnamefont {M.}~\bibnamefont
  {Jalil}}, \bibinfo {author} {\bibfnamefont {S.}~\bibnamefont {Ghee~Tan}},
  \bibinfo {author} {\bibfnamefont {K.}~\bibnamefont {Eason}}, \ and\ \bibinfo
  {author} {\bibfnamefont {J.~F.}\ \bibnamefont {Kong}},\ }\href@noop {}
  {\bibfield  {journal} {\bibinfo  {journal} {Journal of Applied Physics}\
  }\textbf {\bibinfo {volume} {115}} (\bibinfo {year}
  {2014}{\natexlab{a}})}\BibitemShut {NoStop}%
\bibitem [{\citenamefont {Koretsune}\ \emph {et~al.}(2015)\citenamefont
  {Koretsune}, \citenamefont {Nagaosa},\ and\ \citenamefont
  {Arita}}]{koretsune2015control}%
  \BibitemOpen
  \bibfield  {author} {\bibinfo {author} {\bibfnamefont {T.}~\bibnamefont
  {Koretsune}}, \bibinfo {author} {\bibfnamefont {N.}~\bibnamefont {Nagaosa}},
  \ and\ \bibinfo {author} {\bibfnamefont {R.}~\bibnamefont {Arita}},\
  }\href@noop {} {\bibfield  {journal} {\bibinfo  {journal} {Scientific
  reports}\ }\textbf {\bibinfo {volume} {5}},\ \bibinfo {pages} {13302}
  (\bibinfo {year} {2015})}\BibitemShut {NoStop}%
\bibitem [{\citenamefont {Shishir}\ and\ \citenamefont
  {Ferry}(2009)}]{shishir2009velocity}%
  \BibitemOpen
  \bibfield  {author} {\bibinfo {author} {\bibfnamefont {R.}~\bibnamefont
  {Shishir}}\ and\ \bibinfo {author} {\bibfnamefont {D.}~\bibnamefont
  {Ferry}},\ }\href@noop {} {\bibfield  {journal} {\bibinfo  {journal} {Journal
  of Physics: Condensed Matter}\ }\textbf {\bibinfo {volume} {21}},\ \bibinfo
  {pages} {344201} (\bibinfo {year} {2009})}\BibitemShut {NoStop}%
\bibitem [{\citenamefont {Dai}(2015)}]{dai2015euler}%
  \BibitemOpen
  \bibfield  {author} {\bibinfo {author} {\bibfnamefont {J.~S.}\ \bibnamefont
  {Dai}},\ }\href@noop {} {\bibfield  {journal} {\bibinfo  {journal} {Mechanism
  and Machine Theory}\ }\textbf {\bibinfo {volume} {92}},\ \bibinfo {pages}
  {144} (\bibinfo {year} {2015})}\BibitemShut {NoStop}%
\bibitem [{\citenamefont {Rybakov}\ and\ \citenamefont
  {Kiselev}(2019)}]{rybakov2019chiral}%
  \BibitemOpen
  \bibfield  {author} {\bibinfo {author} {\bibfnamefont {F.~N.}\ \bibnamefont
  {Rybakov}}\ and\ \bibinfo {author} {\bibfnamefont {N.~S.}\ \bibnamefont
  {Kiselev}},\ }\href@noop {} {\bibfield  {journal} {\bibinfo  {journal}
  {Physical review B}\ }\textbf {\bibinfo {volume} {99}},\ \bibinfo {pages}
  {064437} (\bibinfo {year} {2019})}\BibitemShut {NoStop}%
\bibitem [{\citenamefont {Jalil}\ \emph
  {et~al.}(2014{\natexlab{b}})\citenamefont {Jalil}, \citenamefont {Tan},
  \citenamefont {Eason},\ and\ \citenamefont {Kong}}]{Jalil2014TopologicalHC}%
  \BibitemOpen
  \bibfield  {author} {\bibinfo {author} {\bibfnamefont {M.~B.~A.}\
  \bibnamefont {Jalil}}, \bibinfo {author} {\bibfnamefont {S.~G.}\ \bibnamefont
  {Tan}}, \bibinfo {author} {\bibfnamefont {K.}~\bibnamefont {Eason}}, \ and\
  \bibinfo {author} {\bibfnamefont {J.}~\bibnamefont {Kong}},\ }\href
  {https://api.semanticscholar.org/CorpusID:122116561} {\bibfield  {journal}
  {\bibinfo  {journal} {Journal of Applied Physics}\ }\textbf {\bibinfo
  {volume} {115}} (\bibinfo {year} {2014}{\natexlab{b}})}\BibitemShut {NoStop}%
\bibitem [{\citenamefont {Zhao}\ \emph {et~al.}(2022)\citenamefont {Zhao},
  \citenamefont {Guo}, \citenamefont {Zeng}, \citenamefont {Shen},
  \citenamefont {Zhang}, \citenamefont {Tomasello}, \citenamefont {Finocchio},
  \citenamefont {Wang},\ and\ \citenamefont {Liang}}]{zhao2022spin}%
  \BibitemOpen
  \bibfield  {author} {\bibinfo {author} {\bibfnamefont {Y.}~\bibnamefont
  {Zhao}}, \bibinfo {author} {\bibfnamefont {D.}~\bibnamefont {Guo}}, \bibinfo
  {author} {\bibfnamefont {Z.}~\bibnamefont {Zeng}}, \bibinfo {author}
  {\bibfnamefont {M.}~\bibnamefont {Shen}}, \bibinfo {author} {\bibfnamefont
  {Y.}~\bibnamefont {Zhang}}, \bibinfo {author} {\bibfnamefont
  {R.}~\bibnamefont {Tomasello}}, \bibinfo {author} {\bibfnamefont
  {G.}~\bibnamefont {Finocchio}}, \bibinfo {author} {\bibfnamefont
  {R.}~\bibnamefont {Wang}}, \ and\ \bibinfo {author} {\bibfnamefont
  {S.}~\bibnamefont {Liang}},\ }\href@noop {} {\bibfield  {journal} {\bibinfo
  {journal} {New Journal of Physics}\ }\textbf {\bibinfo {volume} {24}},\
  \bibinfo {pages} {053053} (\bibinfo {year} {2022})}\BibitemShut {NoStop}%
\bibitem [{\citenamefont {Litzius}\ \emph {et~al.}(2017)\citenamefont
  {Litzius}, \citenamefont {Lemesh}, \citenamefont {Kr{\"u}ger}, \citenamefont
  {Bassirian}, \citenamefont {Caretta}, \citenamefont {Richter}, \citenamefont
  {B{\"u}ttner}, \citenamefont {Sato}, \citenamefont {Tretiakov}, \citenamefont
  {F{\"o}rster} \emph {et~al.}}]{litzius2017skyrmion}%
  \BibitemOpen
  \bibfield  {author} {\bibinfo {author} {\bibfnamefont {K.}~\bibnamefont
  {Litzius}}, \bibinfo {author} {\bibfnamefont {I.}~\bibnamefont {Lemesh}},
  \bibinfo {author} {\bibfnamefont {B.}~\bibnamefont {Kr{\"u}ger}}, \bibinfo
  {author} {\bibfnamefont {P.}~\bibnamefont {Bassirian}}, \bibinfo {author}
  {\bibfnamefont {L.}~\bibnamefont {Caretta}}, \bibinfo {author} {\bibfnamefont
  {K.}~\bibnamefont {Richter}}, \bibinfo {author} {\bibfnamefont
  {F.}~\bibnamefont {B{\"u}ttner}}, \bibinfo {author} {\bibfnamefont
  {K.}~\bibnamefont {Sato}}, \bibinfo {author} {\bibfnamefont {O.~A.}\
  \bibnamefont {Tretiakov}}, \bibinfo {author} {\bibfnamefont {J.}~\bibnamefont
  {F{\"o}rster}},  \emph {et~al.},\ }\href@noop {} {\bibfield  {journal}
  {\bibinfo  {journal} {Nature Physics}\ }\textbf {\bibinfo {volume} {13}},\
  \bibinfo {pages} {170} (\bibinfo {year} {2017})}\BibitemShut {NoStop}%
\bibitem [{\citenamefont {Thiele}(1973)}]{thiele1973steady}%
  \BibitemOpen
  \bibfield  {author} {\bibinfo {author} {\bibfnamefont {A.}~\bibnamefont
  {Thiele}},\ }\href@noop {} {\bibfield  {journal} {\bibinfo  {journal}
  {Physical Review Letters}\ }\textbf {\bibinfo {volume} {30}},\ \bibinfo
  {pages} {230} (\bibinfo {year} {1973})}\BibitemShut {NoStop}%
\end{thebibliography}%

\end{document}